\documentclass[aps,prb,longbibliography,showpacs,floatfix,superscriptaddress,twocolumn]{revtex4-1}

\usepackage{graphicx,color}
\usepackage{amsfonts, amsbsy}
\usepackage[figuresright]{rotating}
\usepackage{amssymb}
\usepackage{amsmath}
\usepackage{bm}
\usepackage{blkarray}
\usepackage{mathtools}
\usepackage{psfrag}
\usepackage{caption}
\usepackage{subcaption}
\usepackage{multirow}
\usepackage{tabularx}
\usepackage{textcomp}
\usepackage{units}
\usepackage{lipsum}
\usepackage{soul}
\usepackage{titlesec}
\usepackage{times}
\usepackage{hyperref}
\usepackage{enumitem}

\DeclareMathAlphabet\mathbfcal{OMS}{cmsy}{b}{n}

\titlespacing*{\section}
{0pt}{1ex plus .25ex}{1ex plus 1ex}
\titlespacing*{\subsection}
{0pt}{1ex plus .25ex}{1ex plus 1ex}
\titlespacing*{\subsubsection}
{0pt}{1ex plus .25ex}{1ex plus 1ex}

\titleformat{\section}
 {\fontsize{10}{15}\bfseries }
  {\thesection}
  {1em}
  {}

\titleformat{\subsection}
  {\bfseries }
  {\thesubsection}
  {2em}
  {}

\titleformat{\subsubsection}
  {\itshape}
  {\thesubsubsection}
  {2em}
  {}

\def\beq{\begin{eqnarray}}
\def\eeq{\end{eqnarray}}

\newcommand{\ket}[1]{\left| #1 \right>} 
\newcommand{\bra}[1]{\left< #1 \right|} 
\let\baraccent=\= 
\renewcommand{\=}[1]{\stackrel{#1}{=}} 
\newcommand{\mc}[1]{\mathcal{ #1}} 

\makeatletter

\titleclass{\subsubsubsection}{straight}[\subsection]

\newcounter{subsubsubsection}[subsubsection]
\renewcommand\thesubsubsubsection{\thesubsubsection.\arabic{subsubsubsection}}

\titleformat{\subsubsubsection}
   {\normalfont\normalsize\itshape \centering}{\thesubsubsubsection}{1em}{}
\titlespacing*{\subsubsubsection}
{0pt}{1ex plus .3ex}{1ex plus .3ex}

\makeatletter
\renewcommand\paragraph{\@startsection{paragraph}{5}{\z@}%
  {3.25ex \@plus1ex \@minus.2ex}%
  {-1em}%
  {\normalfont\normalsize}}
\renewcommand\subparagraph{\@startsection{subparagraph}{6}{\parindent}%
  {3.25ex \@plus1ex \@minus .2ex}%
  {-1em}%
  {\normalfont\normalsize}}
\def\toclevel@subsubsubsection{4}
\def\toclevel@paragraph{5}
\def\toclevel@paragraph{6}
\def\l@subsubsubsection{\@dottedtocline{4}{7em}{4em}}
\def\l@paragraph{\@dottedtocline{5}{10em}{5em}}
\def\l@subparagraph{\@dottedtocline{6}{14em}{6em}}
\makeatother

\begin{document}
\title{Multiplicative topological semimetals}
\author{Adipta Pal}
\affiliation{Max Planck Institute for Chemical Physics of Solids, Nöthnitzer Strasse 40, 01187 Dresden, Germany}
\affiliation{Max Planck Institute for the Physics of Complex Systems, Nöthnitzer Strasse 38, 01187 Dresden, Germany}

\author{Joe H. Winter}
\affiliation{Max Planck Institute for Chemical Physics of Solids, Nöthnitzer Strasse 40, 01187 Dresden, Germany}
\affiliation{Max Planck Institute for the Physics of Complex Systems, Nöthnitzer Strasse 38, 01187 Dresden, Germany}
\affiliation{SUPA, School of Physics and Astronomy, University of St.\ Andrews, North Haugh, St.\ Andrews KY16 9SS, UK}

\author{Ashley M. Cook}
\affiliation{Max Planck Institute for Chemical Physics of Solids, Nöthnitzer Strasse 40, 01187 Dresden, Germany}
\affiliation{Max Planck Institute for the Physics of Complex Systems, Nöthnitzer Strasse 38, 01187 Dresden, Germany}

\begin{abstract}
Exhaustive study of topological semimetal phases of matter in equilibriated electonic systems and myriad extensions has built upon the foundations laid by earlier introduction and study of the Weyl semimetal, with broad applications in topologically-protected quantum computing, spintronics, and optical devices. We extend recent introduction of multiplicative topological phases to find previously-overlooked topological semimetal phases of electronic systems in equilibrium, with minimal symmetry-protection. We show these multiplicative topological semimetal phases exhibit rich and distinctive bulk-boundary correspondence and response signatures that greatly expand understanding of consequences of topology in condensed matter settings, such as the limits on Fermi arc connectivity and structure, and transport signatures such as the chiral anomaly. Our work therefore lays the foundation for extensive future study of multiplicative topological semimetal phases.
 \end{abstract}
\maketitle
\section{Introduction}
Topological semimetals are a vast family~\cite{soluyanov_type-ii_2015, bradlyn2016} of topological phases of matter studied in great depth experimentally~\cite{huang_weyl_2015, PhysRevX.5.031013, lv_observation_2015, xu_discovery_2015, doi:10.1126/science.aaa9297, PhysRevLett.113.027603, liu_stable_2014,
doi:10.1126/science.1245085, neupane_observation_2014} in the search for table-top, quasiparticle realizations of high-energy physics~\cite{PhysRevB.83.205101}. At the simplest level, the topological degeneracies of band structures in these topological semimetal phases are realized quite generically if either time-reversal symmetry~\cite{burkov2011} or inversion symmetry~\cite{halasz2012} are broken. This is the requirement for two-fold topological degeneracies characteristic of the Weyl semimetal phase, although it is desireable to realize such degeneracies in the vicinity of the Fermi level~\cite{RevModPhys.90.015001, PhysRevB.92.161107}, with minimal contributions to the Fermi surface from other electronic states. In such cases, the key signatures of Weyl semimetals are especially prominent, including the distinguishing Fermi arc surface states~\cite{PhysRevX.5.041046, potter2014quantum, Gyenis_2016, doi:10.1126/sciadv.1600709}, and transport signatures associated with the chiral anomaly~\cite{NIELSEN1983389, PhysRevB.88.104412, PhysRevX.4.031035, PhysRevX.5.031023, zhang_signatures_2016, shekhar_extremely_2015, arnold_negative_2016}. Such isolation of Weyl nodes in the vicinity of the Fermi level is also facilitated---and the physics of topological semimetals enriched---by systematic study of these topological phases in compounds with wide-ranging phenomena, including superconductivity, strong spin-orbit coupling, and strong correlations~\cite{GRAF20111, PhysRevB.84.220504, PhysRevB.87.184504, Pan_2013, doi:10.1126/sciadv.1500242, PhysRevLett.67.3310, doi:10.1126/sciadv.aar5832}. Much progress has also been made in identifying other topological semimetals with more complex topological degeneracies~\cite{murakami2007phase, PhysRevB.85.195320, young2012, bradlyn2016} in electronic band structures, protected by a large set of crystalline point group symmetries in combination with additional anti-unitary symmetries such as time-reversal.

The present work returns to the foundations of topological semimetal studies by introducing previously-unidentified topological semimetal phases of matter of electronic systems in equilibrium, which may then be generalized in the same manner as outlined above. We do so by studying the first topological semimetal realizations of multiplicative topological phases, a recently-identified set of topological phases of matter described by Bloch Hamiltonians in an infinitely-large, periodic bulk, which are symmetry-protected tensor products of ``parent'' Bloch Hamiltonians. These multiplicative topological semimetal (MTSM) phases are therefore straightforward constructions described by tensor products of Weyl semimetal Bloch Hamiltonians, yet exhibit rich phenomena distinct from all other known topological semimetals.

We first review the Weyl semimetal phase and its canonical models. We then construct the first examples of multiplicative topological semimetal phases using these past results. The multiplicative topological semimetals are then first characterized in the bulk, and their bulk-boundary correspondence established.

\section{Review of the Weyl semimetal phase and suitable models for constructing multiplicative phases}

The Weyl semimetal is a topologically non-trivial phase of matter characterized by topologically-protected, doubly-degenerate and linearly-dispersing band crossings in the Brillouin zone \cite{wan2011topological}. That is, these band-crossings, known as Weyl points or nodes, cannot be removed from the electronic structure through smooth deformations of the Hamiltonian, but rather only through mutual annihilation of the Weyl nodes, by bringing two nodes of opposite topological charge to the same point in the Brillouin zone to gap out these band-touchings. When the Fermi level intersects only the Weyl nodes of this semimetal phase, their low-energy physics dominates, yielding a variety of intensely-studied exotic phenomena of interest for applications. At the simplest level, the Weyl nodes serve as quasiparticle, table-top realizations of Weyl fermions predicted in high-energy physics. However, they are also a starting point in going well beyond high-energy physics, by tilting the Weyl cone to realize a type-II Weyl semimetal phase \cite{soluyanov2015type}, in which the low-energy physics of the Weyl nodes is not Lorentz-invariant.

Weyl semimetal phases can be realized in effectively non-interacting systems where certain discrete symmetries are \textit{broken} rather than respected, in contrast to many other effectively non-interacting topological phases. They may be derived through symmetry-breaking starting from the Dirac semimetal state \cite{wang2012dirac, young2012dirac}, for instance, (which could be topologically-robust or fine-tuned) by breaking either time-reversal symmetry $\mathcal{T}$ or inversion symmetry $\mathcal{I}$, which pulls the two Weyl nodes comprising the Dirac node away from one another in momentum-space \cite{zyuzin2012weyl}. This phase, characterized by Weyl nodes in the Brillouin zone, is topologically stable so long as Weyl nodes of opposite topological charge do not annihilate one another \cite{morimoto2014weyl}.

$\mc{I}$-breaking Weyl semimetal phases are of tremendous experimental interest, but are described by Bloch Hamiltonian models with four bands at minimum. A more natural starting point in deriving multiplicative topological semimetal phases is instead to use the minimal Weyl semimetal Bloch Hamiltonian achieved by breaking $\mc{T}$, which possesses only two bands. Such two-band models for the Weyl semimetal correspond to the non-trivial homotopy group $\pi_3(S^2)$ and, similarly to the two-band Chern and Hopf insulators~\cite{moore2008} and the two-band Kitaev chain model ~\cite{Kitaev_2001}, may be combined using known constructions to form a multiplicative counterpart of the Weyl semimetal phase, the multiplicative Weyl semimetal phase (MWSM).

We therefore consider a well-established two-band Bloch Hamiltonian previously used in study of Weyl nodes, with various instances of this model serving as the parents of the MWSM.

\begin{equation}
\begin{split}
\mc{H}_{WSM}(\boldsymbol{k})=&t_1\sin k_x\tau^x+t_2\sin k_y\tau^y\\
&+t_3(2+\gamma-\cos k_x-\cos k_y-\cos k_z)\tau^z.
\end{split}
\end{equation}
where the $\tau^j$ ($j=x,y,z$) are the Pauli matrices in the orbital basis. The two band spectrum,
\begin{equation}
\begin{split}
&E(\boldsymbol{k})=\pm \sqrt{t_1^2\sin^2k_x+t_2^2\sin^2k_y+\epsilon(\boldsymbol{k})^2},\\
& \epsilon(\boldsymbol{k})=t_3(2+\gamma-\cos k_x-\cos k_y-\cos k_z),
\end{split}
\end{equation}
has two gapless nodes at $\boldsymbol{k}=(0,0,\pm k_0)$, for $\cos k_0=\gamma$. We refer to these as the Weyl nodes. The equation of motion for Bloch electrons in the $\boldsymbol{k}$-space in the presence of Berry curvature is represented by $\dot{\mathbf{r}}=\mathbf{v}_{\boldsymbol{k}}+\dot{\boldsymbol{k}}\times\mathbf{F}(\boldsymbol{k})$. For the equation of motion to remain invariant under $\mathcal{T}$-symmetry, one must have the equality, $\mathbf{F}(\boldsymbol{k})=-\mathbf{F}(-\boldsymbol{k})$. The breaking of $\mathcal{T}$-symmetry, then involves a minimum of two Weyl nodes with opposite Berry curvature at opposite momenta. Therefore, close to the Weyl nodes, we have,
\begin{equation}
\mc{H}_\pm(\boldsymbol{k})=\pm t_1k_x\tau^x+t_2k_y\tau^y\pm t_3\sin k_0 k_z\tau^z,
\end{equation}
which in turn corresponds to the Berry curvatures,
\begin{equation}
\mathbf{F}^\pm(\boldsymbol{k})|_{0,0,\pm k_0}=\pm \frac{t_1t_2t_3\sin k_0}{2[t_1k_x^2+t_2k_y^2+(t_3\sin k_0)^2k_z^2]^{3/2}}(k_x,k_y,k_z).
\end{equation}
The Chern number of the lower-energy band for the range, $k_x=0$, $k_y=0$ and $k_z\in (-k_0,k_0)$ is $C = \pm 1$ depending on the direction of the magnetic field corresponding to the monopoles at the two Weyl points. The Weyl nodes are involved with exotic boundary states at surfaces perpendicular to the $z$-axis, called the Fermi Arc surface states. For the case where the surfaces are open in the $x$-direction, the surface dispersion is given by,
\begin{equation}
E(k_y) = \pm t_2\sin k_y,
\end{equation}
and the arc-states,
$$
\Psi(x,k_y,k_z) = e^{+ik_yy+ik_zz}(e^{-\lambda_1x}-e^{-\lambda_2x})
\frac{1}{\sqrt{2}}
\begin{pmatrix}
1\\
\pm i
\end{pmatrix}
.
$$
In the $k$-space, this includes all contours $\cos k_y+\cos k_z >1+\cos k_0$.

\section{Multiplicative Weyl Semimetal (MWSM) in the bulk}
A protocol for constructing the child Hamiltonian for the MWSM, $\mc{H}_c$ derived from $\mc{H}_{p1}$ and $\mc{H}_{p2}$ as first reported in Cook and Moore~\cite{cook2022mult}, is given as follows. Given two two-band Bloch Hamiltonians $\mc{H}_{p1}$ and $\mc{H}_{p2}$ written in a general form, with momentum-dependence suppressed, as
\begin{equation}
\begin{split}
\mc{H}_{p1}=
\begin{pmatrix}
a & b\\
c & d
\end{pmatrix}
;\quad \mc{H}_{p2}=
\begin{pmatrix}
\alpha & \beta\\
\gamma & \delta
\end{pmatrix},\\
\end{split}
\end{equation}

the multiplicative child Bloch Hamiltonian constructed from these two parents can be written as $\mc{H}^c_{12}$, where
\begin{equation}
    \mc{H}^c_{12}=
\begin{pmatrix}
a\delta & -a\gamma & b\delta & -b\gamma\\
-a\beta & a\alpha & -b\beta & b\alpha\\
c\delta & -c\gamma & d\delta & -d\gamma\\
-c\beta & c\alpha & -d\beta & d\alpha
\end{pmatrix}.
\end{equation}

Expressing the two-band parent Bloch Hamiltonians $\mc{H}_{p1}(\boldsymbol{k})$ and $\mc{H}_{p2}(\boldsymbol{k})$ more compactly as the following,
\begin{equation}
\begin{split}
&\mc{H}_{p1}(\boldsymbol{k})=\mathbf{d}_1 (\boldsymbol{k})\cdot \boldsymbol{\tau};\quad \mc{H}_{p2}(\boldsymbol{k})=\mathbf{d}_2 (\boldsymbol{k}) \cdot \boldsymbol{\sigma},
\end{split}
\end{equation}
where $\mathbf{d}_1(\boldsymbol{k})$ and $\mathbf{d}_2(\boldsymbol{k})$ are momentum-dependent, three-component vectors of scalar functions, and each of $\boldsymbol{\sigma}$ and $\boldsymbol{\tau}$ is the vector of Pauli matrices, the multiplicative child Hamiltonian may more compactly be written as,
\begin{equation}
\begin{split}
& \mc{H}^c_{12}(\boldsymbol{k})=(d_{11},d_{21},d_{31})\cdot\boldsymbol{\tau}\otimes (-d_{12},d_{22},-d_{32})\cdot \boldsymbol{\sigma},
\end{split}
\end{equation}
to highlight the tensor product structure of the child Hamiltonian, which can be symmetry-protected as discussed in earlier work by Cook and Moore on multiplicative topological phases, and therefore can describe phases of matter, even in the presence of additional bands~\cite{cook2022mult}.

The tensor-product structure guarantees that the energy spectrum of the child Hamiltonian is a product of the energy spectrum of $\mc{H}_{p1}(\boldsymbol{k})$, $E_{p1}(\boldsymbol{k})$, and of $\mc{H}_{p2}(\boldsymbol{k})$, $E_{p2}(\boldsymbol{k})$, respectively,
\begin{equation}
E^c_{12}(\boldsymbol{k})= \pm E_{p1}(\boldsymbol{k})E_{p2}(\boldsymbol{k}).
\label{eq2xdegen}
\end{equation}
This implies that bands of the child Hamiltonian dispersion are \textit{at least} doubly degenerate everywhere in the bulk Brillouin zone.

We will consider two cases in this work: (1) the Weyl node separation of each parent is along one axis in the Brillouin zone, and (2) the axis along which Weyl nodes are separated in one parent is perpendicular to the axis along which Weyl nodes are separated in the other parent. Spectral and magneto-transport properties differ significantly between these two cases, as we will show, demonstrating the richness of MTSM phases of matter.

\subsection{Multiplicative Weyl Semimetal - parallel axis parents}\label{MWSMpllintro}
The construction of the MWSM for both parents along the same axis is derived from two parent WSMs. As an example, we consider the following parents and the resulting child:
\begin{subequations}
\begin{equation}
\begin{split}
\mc{H}_{p1}(\boldsymbol{k})=& t_{11}\sin k_x\tau^x+t_{21}\sin k_y\tau^y\\
& +t_{31}(2+\gamma_1-\cos k_x-\cos k_y-\cos k_z)\tau^z,\\
\end{split}
\label{Hws1par}
\end{equation}
\begin{equation}
\begin{split}
\mc{H}_{p2}(\boldsymbol{k})=&t_{12}\sin k_x\sigma^x+t_{22}\sin k_y\sigma^y\\
&+t_{32}(2+\gamma_2-\cos k_x-\cos k_y-\cos k_z)\sigma^z,
\end{split}
\label{Hws2par}
\end{equation}
\begin{equation}
\begin{split}
\mc{H}_c(\boldsymbol{k})=& [t_{11}\sin k_x\tau^x+t_{21}\sin k_y\tau^y\\
& +t_{31}(2+\gamma_1-\cos k_x-\cos k_y-\cos k_z)\tau^z]\\
& \otimes [-t_{12}\sin k_x\sigma^x+t_{22}\sin k_y\sigma^y\\
&-t_{32}(2+\gamma_2-\cos k_x-\cos k_y-\cos k_z)\sigma^z].
\end{split}
\label{Hmwspll}
\end{equation}
\end{subequations}

\begin{figure}[h]
    \centering
    \includegraphics[scale=0.6]{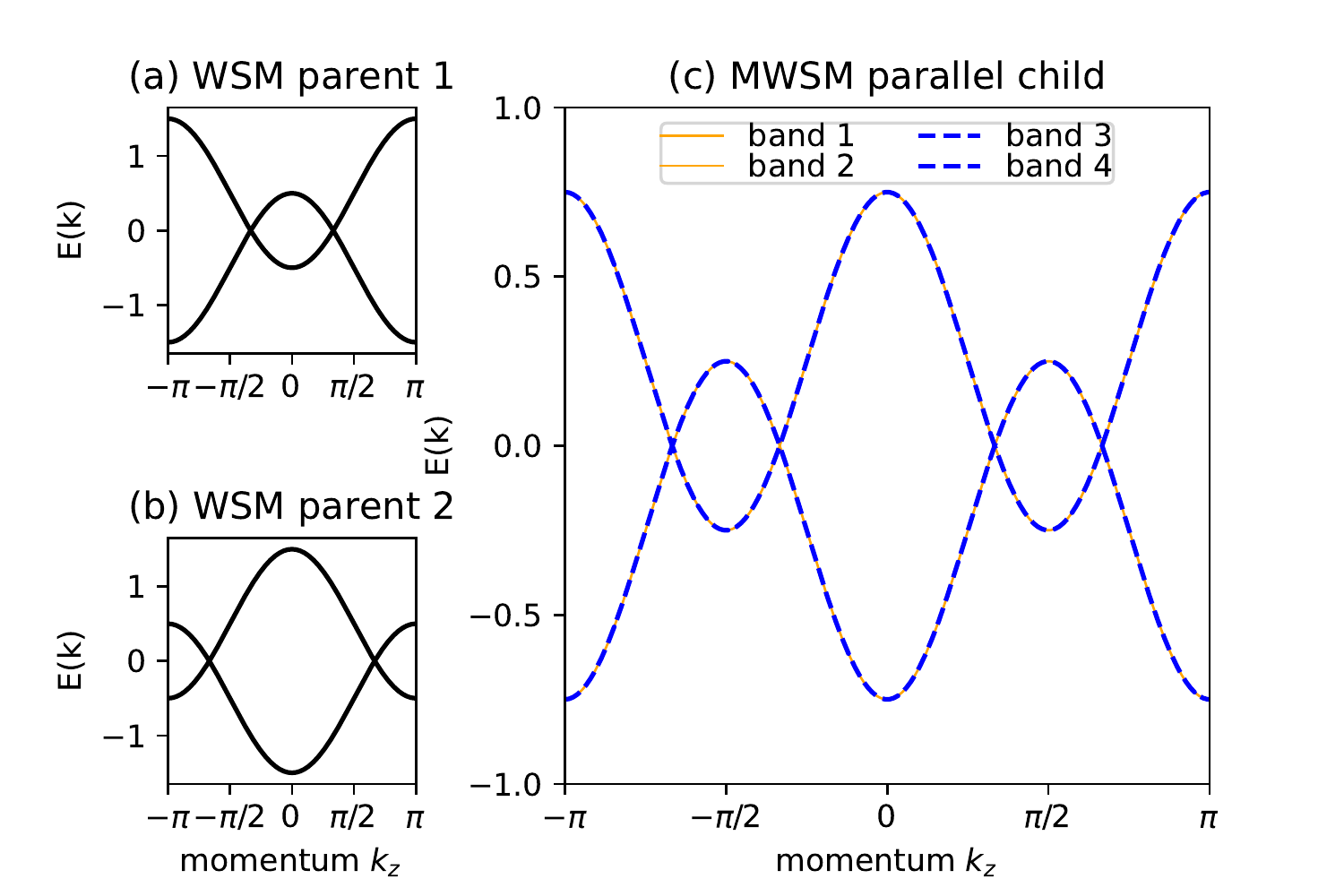}
    \caption{Dispersion $E(\boldsymbol{k})$ for (a) WSM Parent Hamiltonian with $\gamma_1=0.5$ along $k_z$ and $t_{11}=t_{21}=t_{31}=1$, (b) WSM Parent Hamiltonian with $\gamma_2=-0.5$ along $k_z$ and $t_{12}=t_{22}=t_{32}=1$, and (c) the resulting MWSM parallel Child Hamiltonian along $k_z$.}
    \label{MWSMpll05m05}
\end{figure}

Each parent Hamiltonian realizes Weyl nodes at $\boldsymbol{k} = \left(0,0, \cos^{-1}\gamma_i \right)$ when $-1<\gamma_i<1$, $(i=1,2)$. Examples of such topologically non-trivial dispersion are shown in Fig.~\ref{MWSMpll05m05} (a) and (b), respectively.

From these parent Hamiltonian dispersions, we can find the dispersion of the child. As given in Eq.~\ref{eq2xdegen}, the bulk spectrum is doubly degenerate and determined by the spectra of the parent $1$, $E_{p1}(\boldsymbol{k})$, and parent $2$, $E_{p2}(\boldsymbol{k})$, respectively, which take the following forms:
\begin{equation}
\begin{split}
& E_{p1}(\boldsymbol{k})=[t_{11}^2\sin^2 k_x+t_{21}^2\sin^2 k_y+\epsilon_1(\boldsymbol{k})^2]^{1/2},\\
& E_{p2}(\boldsymbol{k})=[t_{12}^2\sin^2 k_x+t_{22}^2\sin^2 k_y+\epsilon_2(\boldsymbol{k})^2]^{1/2},
\end{split}
\label{mwsmparbulkdisp}
\end{equation}
where $\epsilon_{1/2}(\boldsymbol{k})=t_{31/2}(2+\gamma_{1/2}-\cos k_x-\cos k_y-\cos k_z)$. \\
For the sake of convenience, we refer to the MWSM with Weyl node separation for each parent along the same axis in the Brillouin zone (as in the case of parents given by Eq.~\ref{Hws1par} and Eq.~\ref{Hws2par}) as MWSM$||$. For the MWSM$||$ bulk spectrum given by Eq.~\ref{eq2xdegen} and Eq.~\ref{mwsmparbulkdisp}, gapless points occur at the positions in the Brillouin zone where gapless points are present for the parent systems. As $\gamma_1$ and $\gamma_2$ control separation of the Weyl nodes in the Brillouin zone for the parents, they play a major role in determining the number of nodes, the location of the nodes, and the polynomial order of the nodes in the Brillouin zone for the child. When $\gamma_1=\gamma_2$, for instance, we have two gapless points but the dispersion near the nodes is quadratic. In contrast, for $\gamma_1\neq \gamma_2$ as for parents depicted in Fig.~\ref{MWSMpll05m05} (a) and (b), the child MWSM$||$ has four nodes, and bands disperse linearly in the vicinity of the nodes, as shown in Fig.~\ref{MWSMpll05m05} (c). Each node is four-fold degenerate.

While such degeneracy naively suggests Dirac nodes or Weyl nodes of higher charge, the multiplicative nodes are distinct in a number of ways. To examine this difference, we look at the child Hamiltonian in the vicinity of each multiplicative node for the case $-1<\gamma_1\neq\gamma_2<1$. From the tensor product structure, it easy to check that $\frac{\partial E_{\pm}}{\partial k_i}=const.$ which implies that the dispersion is linear at each of the gapless nodes of the MWSM. Therefore the possibility of a higher order Weyl node is nullified. The position of each of the multiplicative nodes are determined by the nodes in the respective parents. We refer to $(0,0,\pm k_{01})$ as the Weyl node positions derived from the first parent, and $(0,0,\pm k_{02})$ as the Weyl node positions derived from the second parent. Here $\gamma_i=\cos k_{0i}$, $(i=1,2)$. If the gapless point is $(0,0,k_{02})$, then we define MWSM$||$ in the vicinity as $\mc{H}^c_{||,2}$, and,
\begin{equation}
\mc{H}^c_{||,2} = t_{31}(\gamma_1-\gamma_2)\tau^z(-t_{12}k_x\sigma^x+t_{22}k_y\sigma^y-t_{32}\sin k_{02}\bar{k}_{z,2}\sigma^z),
\label{MWSMpllnode2dirac}
\end{equation}
where $\bar{k}_{z,2}=(k_z-k_{02})$. Surprisingly, this looks like a Dirac semimetal Hamiltonian, whose Dirac node has been shifted in $\boldsymbol{k}$-space. Since it is no longer at the origin, the time-reversal symmetry is broken. For the other node, $\gamma_1=\cos k_{01}$ for $(0,0,k_{01})$, we define the multiplicative Hamiltonian in the vicinity as $\mc{H}^c_{||,1}$, so that,
\begin{equation}
\mc{H}^c_{||,1} = (t_{11}k_x\tau^x+t_{21}k_y\tau^y+t_{31}\sin k_{01}\bar{k}_{z,1}\tau^z)t_{32}(\gamma_1-\gamma_2)\sigma^z,
\label{MWSMpllnode1dirac}
\end{equation}
where $\bar{k}_{z,1}d=(k_z-k_{01})$ and contains off-diagonal terms for the block Hamiltonian. But again, it is possible to perform a similarity transformation on this Hamiltonian, in the form $U = R^{-1}_\tau(\theta,\phi)\otimes R_\sigma(\theta,\phi)$, so that we get another `shifted' Dirac semimetal type Hamiltonian,
\begin{equation}
\bar{H}^c_{||,1} = t_{32}(\gamma_1-\gamma_2)\tau^z(t_{11}k_x\sigma^x+t_{21}k_y\sigma^y+t_{31}\sin{k_{01}}\bar{k}_{z,1}\sigma^z).
\end{equation}
Again, the shift from the origin breaks the time-reversal symmetry of the original Dirac semimetal. It is therefore appropriate to refer to the MWSM$||$ as possessing degeneracies consisting of Weyl nodes, rather than possessing Dirac nodes, and exhibit strikingly different physics as a result.

\subsection{MWSM - perpendicular axis parents}\label{MWSMperpbulk}
Before characterizing bulk-boundary correspondence and transport signatures of MTSMs, we explore further richness of multiplicative constructions by considering cases where parent Weyl nodes are separated along orthogonal axes in $\boldsymbol{k}$-space.
As a specific case, we choose parent Hamiltonians such that the first parent has Weyl node separation along the y-axis, while the second one has Weyl node separation along the z-axis,
\begin{subequations}
\begin{equation}
\begin{split}
\mc{H}_{p1}(\boldsymbol{k}) =& t_{11}\sin k_x\tau^x+t_{21}\sin k_z\tau^y\\
&+t_{31}(2+\gamma_1-\sum_i\cos k_i)\tau^z,
\end{split}
\end{equation}
\begin{equation}
\begin{split}
\mc{H}_{p2}(\boldsymbol{k})=& t_{12}\sin k_x\sigma^x+t_{22}\sin k_y\sigma^y\\
&+t_{32}(2+\gamma_2-\sum_i\cos k_i)\sigma^z.
\end{split}
\end{equation}
\end{subequations}



\begin{figure}[htb]
    \centering
    \includegraphics[scale=0.7]{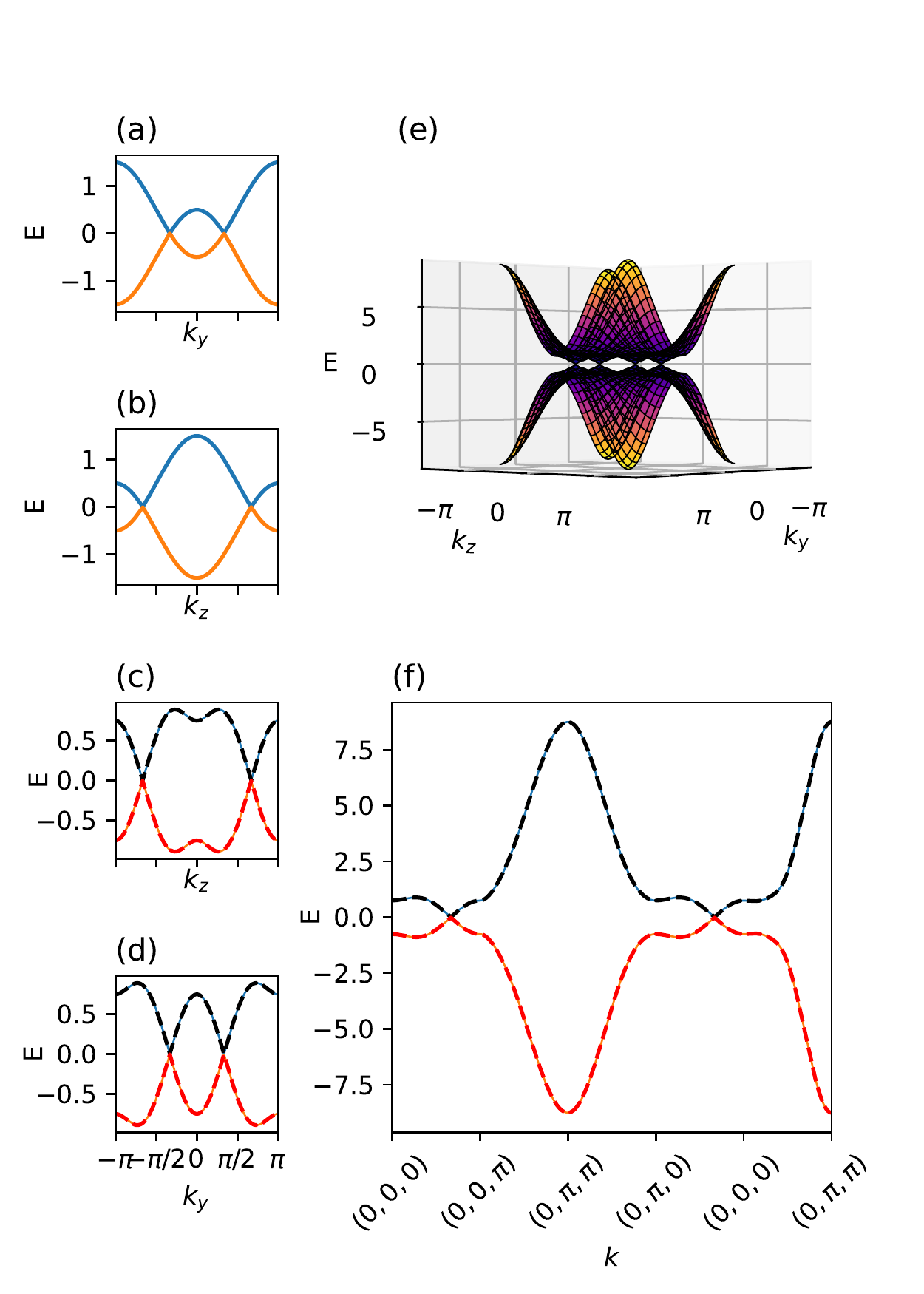}
    \caption{Dispersion $E(k)$ ($t_{11}=t_{12}=1$, $t_{21}=t_{22}=1$, $t_{31}=t_{32}=1$) for (a) WSM Parent Hamiltonian with $\gamma_1=0.5$ along $k_y$, (b) WSM Parent Hamiltonian with $\gamma_2=-0.5$ along $k_z$ and the resulting MWSM perpendicular Child Hamiltonian along (c) $k_z$ and (d) $k_z$. The energy dispersion plotted along both $k_y$ and $k_z$ is shown in (e) and the dispersion along a high-symmetry path in the first quadrant of the two-dimensional (2d) BZ is shown in (f). Inversion symmetry relates the nodes in the first quadrant to those in the other quadrants, giving rise to four gapless nodes in the 2d BZ.} \label{MSWMperpspaghetti}
\end{figure}

Again the bulk spectrum is derived from the tensor product structure,
\begin{equation}
\begin{split}
&E_{p1}(\boldsymbol{k}) = [t_{11}^2\sin^2k_x+t_{21}^2\sin^2 k_z+\epsilon_1^2(\boldsymbol{k})]^{1/2},\\
&E_{p2}(\boldsymbol{k}) = [t_{12}^2\sin^2k_x+t_{22}^2\sin^2k_y+\epsilon^2_2(\boldsymbol{k})]^{1/2},\\
&E^c_\perp{\boldsymbol{k}} = \pm E_{p1}(\boldsymbol{k})E_{p2}(\boldsymbol{k}),
\end{split}
\end{equation}
where $\epsilon_{1/2}(\boldsymbol{k})=t_{31/32}(2+\gamma_{1/2}-\cos k_x-\cos k_y-\cos k_z)$. Examples of parent and child dispersion in this case are shown in Fig.~\ref{MSWMperpspaghetti} for the values, $\gamma_1=0.5$ and $\gamma_2=-0.5$.

We gain greater understanding of the multiplicative structure in this case by examining the low-energy expansion of the Child Hamiltonian in the vicinity of its nodes. Taylor expanding up to linear order around the point, $(0,k_{0,1},0)$ for $\gamma_1=\cos k_{0,1}$, one gets,
\begin{equation}
\begin{split}
\mc{H}^c_{\perp,1}(\boldsymbol{k}) =& (t_{11}k_x\tau^x+t_{21}k_z\tau^y+t_{31}\sin k_{0,1}\bar{k}_{y,1}\tau^z)\\
&\otimes (t_{22}\sin k_{0,1}\sigma^y-t_{32}(\gamma_2-\gamma_1)\sigma^z).
\end{split}
\label{MWSMperpliny}
\end{equation}
Similarly, expanding around $(0,0,k_{0,2})$ for $\gamma_2=\cos k_{0,2}$, we get,
\begin{equation}
\begin{split}
\mc{H}^c_{\perp,2}(\boldsymbol{k}) =& (t_{21}\sin k_{0,2}\tau^y+t_{31}(\gamma_1-\gamma_2)\tau^z)\\
&\otimes (-t_{12}k_x\sigma^x+t_{22}k_y\sigma^y-t_{32}\sin k_{0,2}\bar{k}_{z,2}\sigma^z).
\end{split}
\label{MWSMperplinz}
\end{equation}

One notices that $\mc{H}^c_{\perp,2}(\boldsymbol{k})$ is equivalent to a DSM when $\gamma_1=\gamma_2$.

\subsection{Discrete Symmetries of the MWSM}
The discrete symmetries satisfied by the parent WSMs include invariance under particle-hole conjugation given by $\mathcal{P}=\sigma^x\kappa$, such that the Hamiltonian satisfies,
$$
\sigma^x\mc{H}^*_{1/2}(\boldsymbol{k})\sigma^x=-\mc{H}_{1/2}(-\boldsymbol{k}),
$$
and invariance under spatial inversion given by $\mc{I}=\sigma^z$, such that the Hamiltonian satisfies,
$$
\sigma^z\mc{H}_{1/2}(\boldsymbol{k})\sigma^z=\mc{H}_{1/2}(-\boldsymbol{k}).
$$
The MWSM$||$ or $\perp$ child systems are instead invariant under time reversal given by $\mc{T}=i\tau^x\sigma^x\kappa$ corresponding to the transformation,
$$
\tau^x\sigma^x\mc{H}^*_{c}(\boldsymbol{k})\tau^x\sigma^x = \mc{H}_{c}(-\boldsymbol{k}).
$$
They are also invariant under spatial inversion given by $\mc{I}=\tau^z\sigma^z$, corresponding to the transformation,
$$
\tau^z\sigma^z\mc{H}_c(\boldsymbol{k})\tau^z\sigma^z=\mc{H}_c(-\boldsymbol{k}).
$$
The MWSM should then satisfy the symmetry, $\mc{TI}$, which may also protect the Dirac semi-metal phase. Indeed, in some cases, the Dirac Hamiltonian for the MWSM near the nodes is reminiscent of the corresponding low-energy Hamiltonian for a Dirac semi-metal. This invariance of the multiplicative bulk Hamiltonian under products of transformations, which leave each parent Hamiltonian invariant, is expected given the multiplicative dependence of the child on the parents.

\subsection{Bulk characterization of topology with Wilson loops}
\begin{figure}[htb!]
    \centering
    \begin{subfigure}{0.49\textwidth}
    \includegraphics[width=0.95\textwidth]{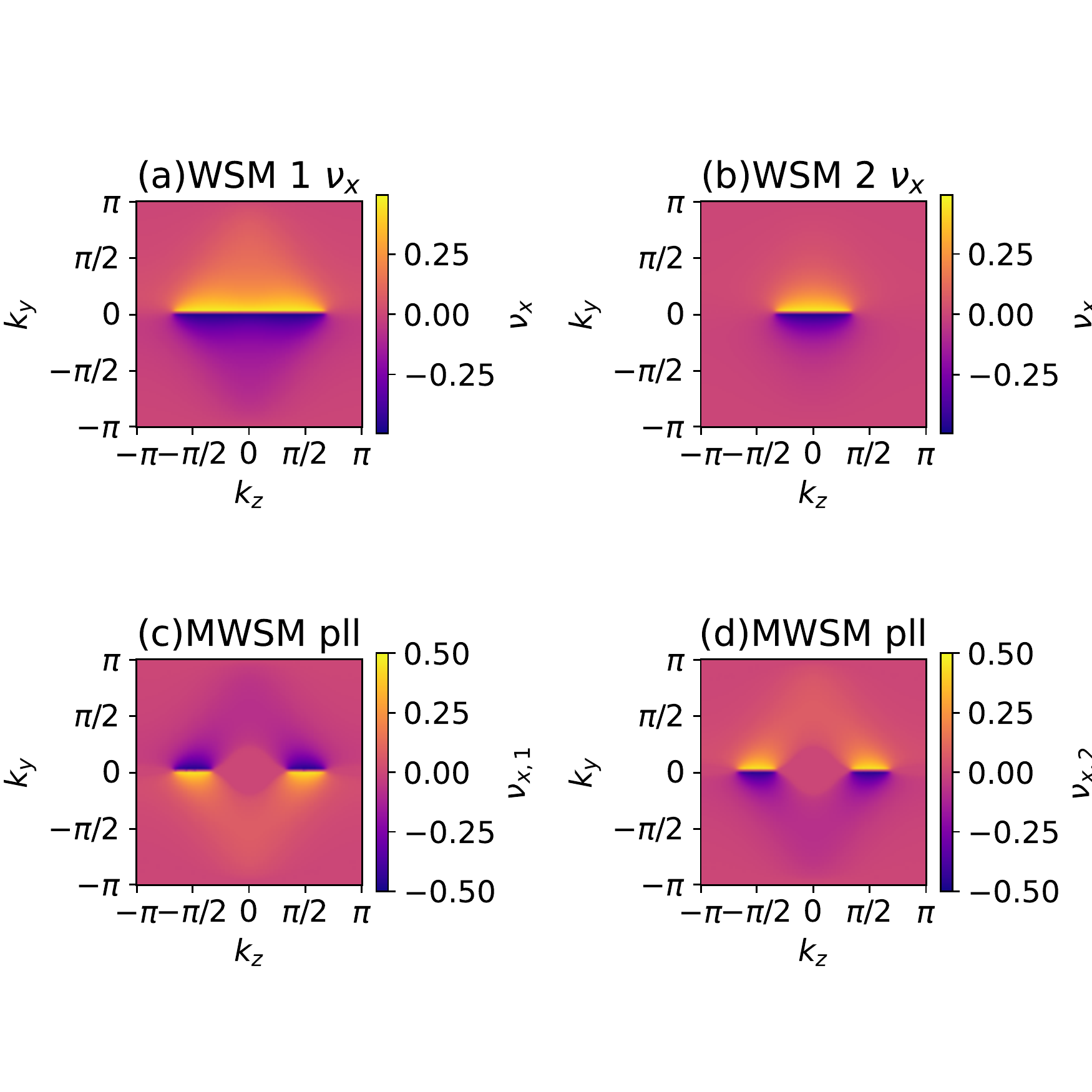}
    \caption{Each parent has Weyl node `dipole' oriented in $+\hat{z}$ direction.}
    \end{subfigure}
    ~
    \begin{subfigure}{0.49\textwidth}
    \includegraphics[width=0.95\textwidth]{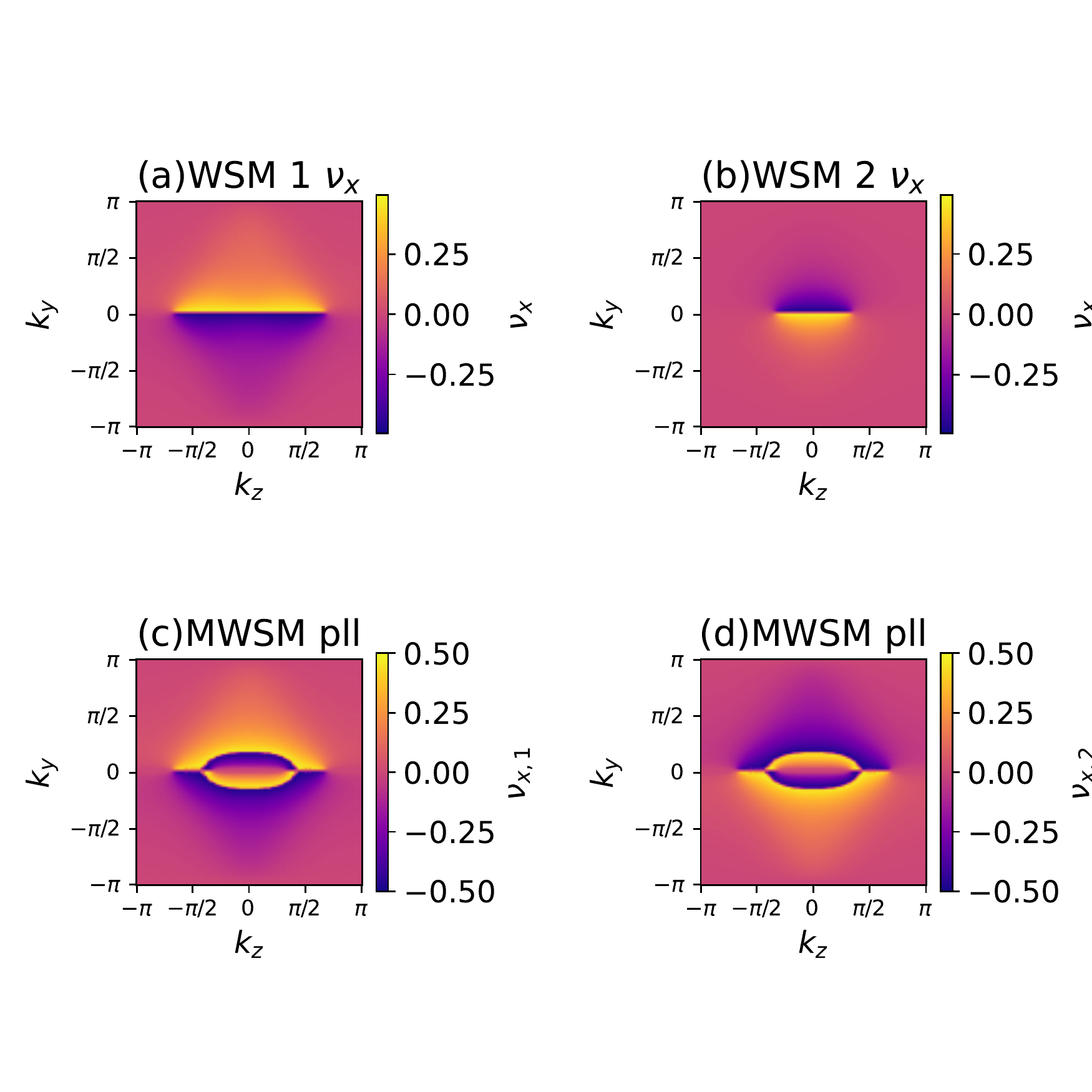}
    \caption{Parent 1 with dipole oriented along $+\hat{z}$ direction and parent 2 with dipole oriented in $-\hat{z}$ direction.}
    \end{subfigure}
    \caption{Wannier spectra in MWSM parallel for two filled bands derived from Wilson loop around $k_x$ for parent 1 with $\gamma_1=-0.5$ and parent 2 with $\gamma_2=0.5$. The upper row (a) and the lower row (b) have opposite orientation of the Weyl node 'dipole' for parent 2. Corresponding Wannier spectra of the MWSM for the lowest-energy and second-lowest in energy occupied bands is shown in (c) and (d), respectively.}
    \label{WLpll}
\end{figure}

\begin{figure}[htb!]
    \centering
    \begin{subfigure}{0.49\textwidth}
    \includegraphics[width=0.95\textwidth]{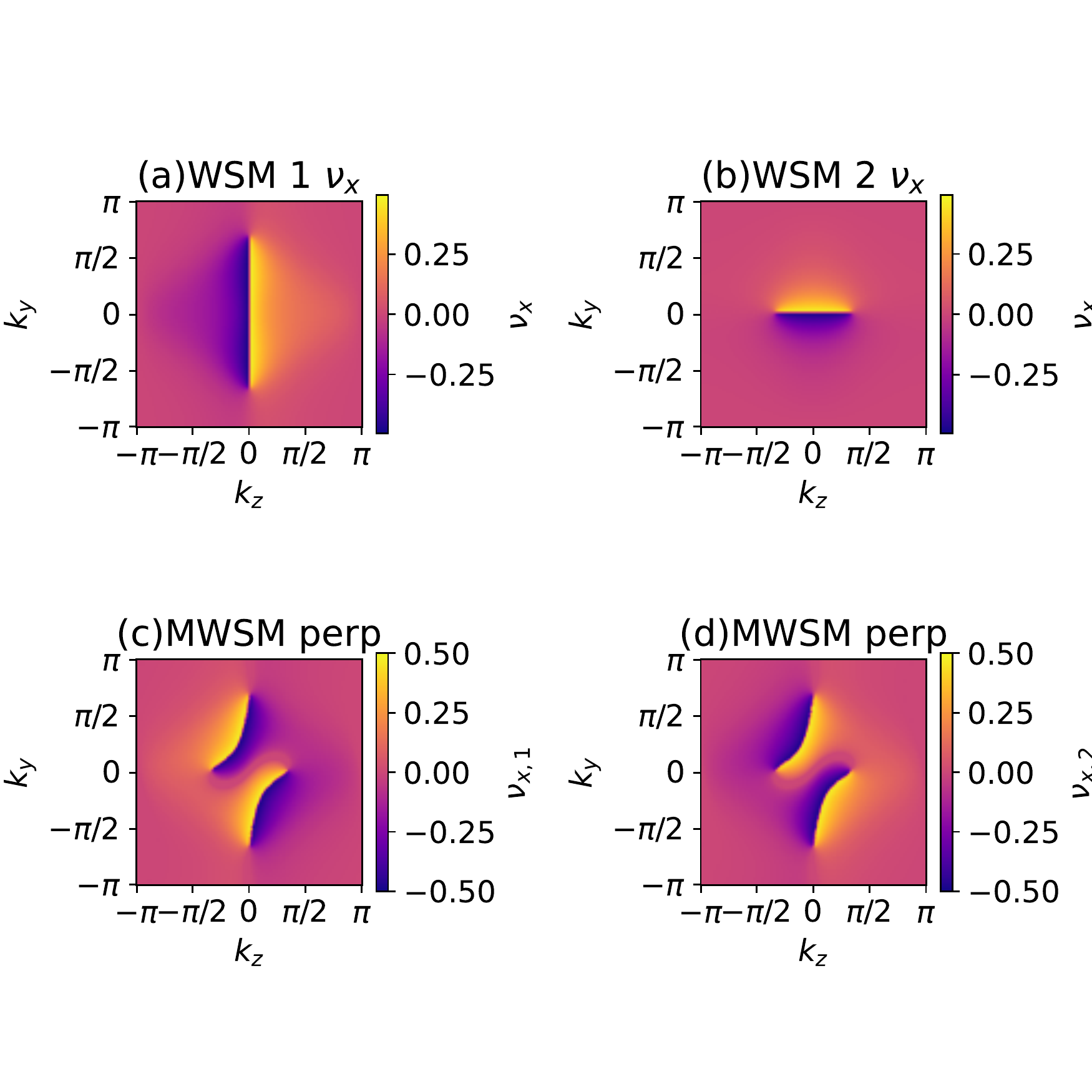}
    \caption{Parent 1 has Weyl node `dipole' oriented along $+\hat{y}$ direction and parent 2 along $-\hat{z}$ direction.}
    \end{subfigure}
    ~
    \begin{subfigure}{0.49\textwidth}
    \includegraphics[width=0.95\textwidth]{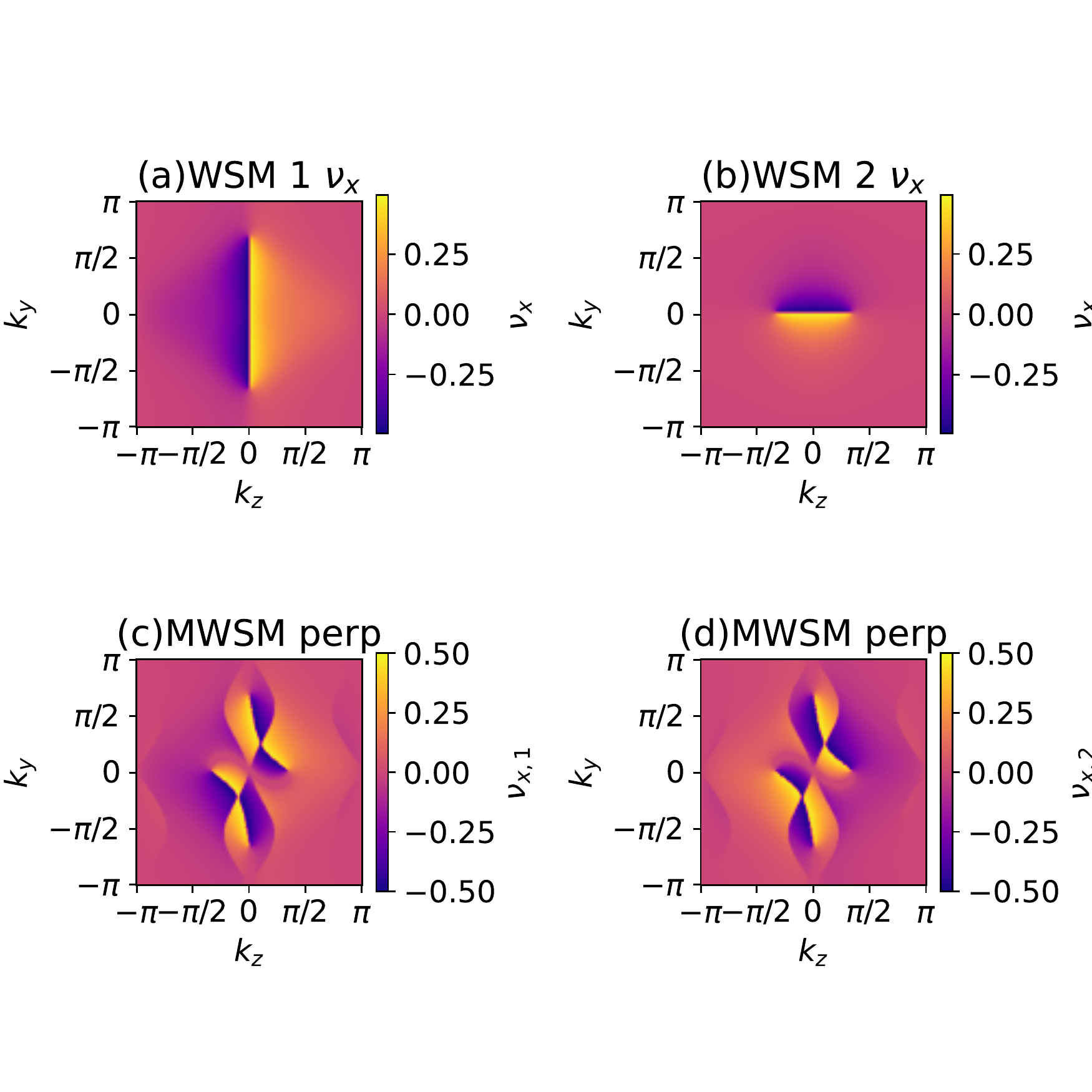}
    \caption{Parent 1 has Weyl node 'dipole' oriented in $+hat{y}$ direction and parent 2 Weyl node dipole oriented in $-\hat{z}$ direction.}
    \end{subfigure}
    \caption{Wannier spectra in MWSM perpendicular for two filled bands derived from Wilson loop around $k_x$ for parent 1 with $\gamma_1=-0.5$ and parent 2 with $\gamma_2=0.5$. The upper row (a) and the lower row (b) have opposite orientation of the Weyl node 'dipole' for parent 2. Corresponding Wannier spectra of the MWSM for the lowest-energy and second-lowest in energy occupied bands is shown in (c) and (d), respectively.}
    \label{WLperp}
\end{figure}

As calculated in Supplementary section \ref{Wilsonloopsup}, the Berry connection for the MWSM is given as
\begin{equation}
\mathbf{A}=(A_{1,k_x}-A_{2,k_x},A_{1,k_y}-A_{2,k_y},A_{1,k_z}-A_{2,k_z}),
\label{BerryM1}
\end{equation}
where $A_{j,l}=(i\bra{+_j}\partial_l\ket{+_j},i\bra{-_j}\partial_l\ket{-_j})$. Using this expression for the Berry connection, we compute Wilson loops and associated Wannier spectra by integrating over $k_x$ for a given $k_y$, as detailed in Alexandradinata~\emph{et al}~\cite{WLpaper}. In the parallel case illustrated in Fig. \ref{WLpll}(a), the Wannier spectra derived from Wilson loop calculations show that only in regions where only one of the parent phases is non-trivial do we get non-trivial Wannier spectra distinguished by $\pi$ values for Wannier charge centers. However, the Wannier spectra in the region where each parent is topological appears trivial, given the dependence of child Wannier spectra on parent Wannier spectra distinctive of multiplicative topological phases. We have referred to a pair of Weyl nodes of equal and opposite topological charge as a `dipole'. We observe, that the orientation of this dipole due to the two constituent parents is important, as anti-parallel dipoles, as depicted in Fig. \ref{WLpll}(b), show non-trivial Wilson loop eigenvalues in a region in the 2d BZ where neither of the parent systems have non-trivial topological character. Analogous results for the MWSM$\perp$ are shown in Fig.~\ref{WLperp}, although the Wannier spectrum structure is far richer than in the parallel case.\\

\break

\section{MWSM with open-boundary conditions--}
\subsubsection{Slab spectra of MWSM:}

An important aspect of WSM physics is its distinctive bulk-boundary correspondence: Weyl nodes in the three-dimensional bulk Brillouin zone serve as termination points of topologically-protected boundary states known as Fermi arcs when projected to a slab Brillouin zone corresponding to open boundary conditions in one direction. We expect analogous topologically-protected surface states in MTSMs and explore possible realizations of these Fermi arc states in this section.

\begin{figure}[htb!]
\centering
    \includegraphics[scale=0.65]{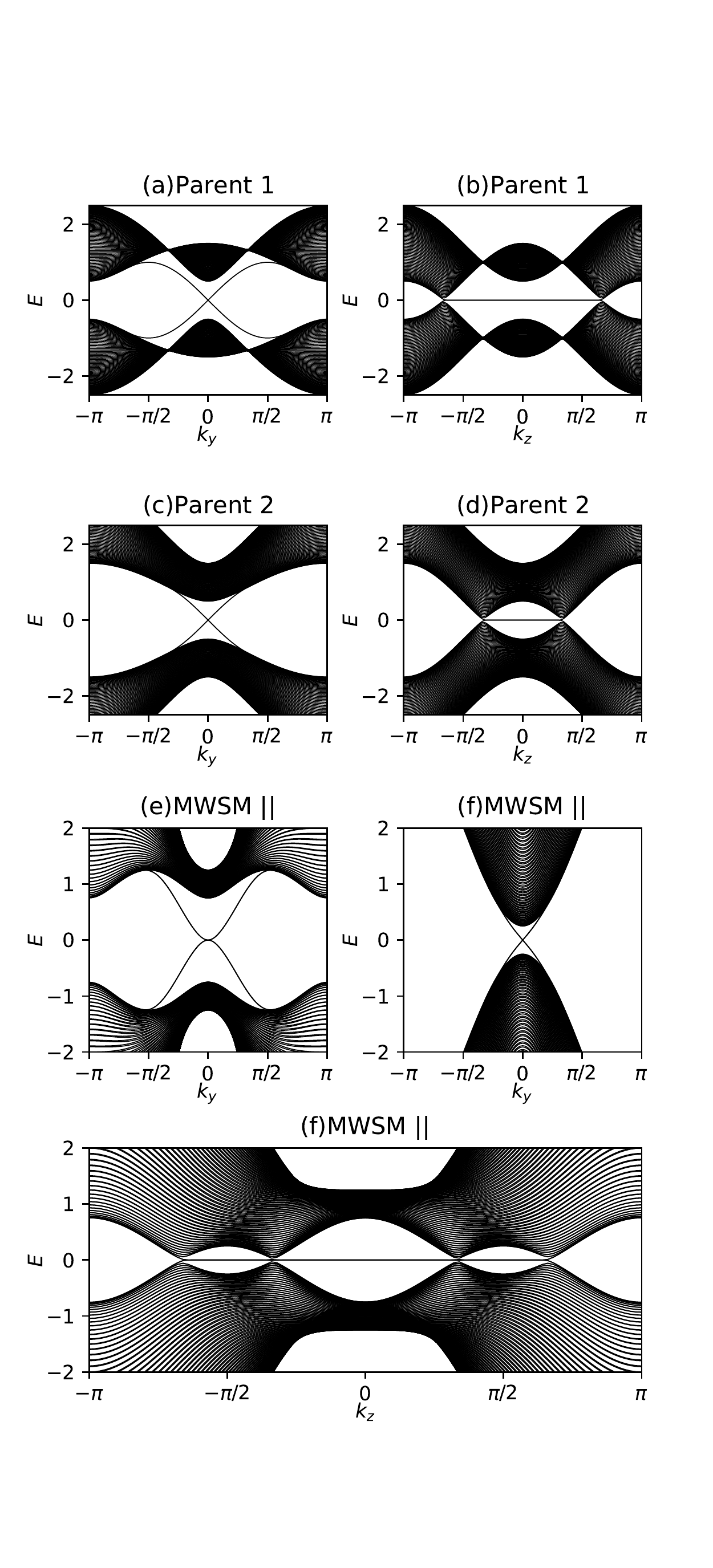}
    \caption{Finite slab spectra(in x-direction, $L_x=80$) along $k_y$ ($k_z=0$) and $k_z$($k_y=0$) respectively for (a,b) WSM with $\gamma_1=-0.5$, (c,d) WSM with $\gamma_2=0.5$. In (e,f) the slab spectra($L_x=80$) $E$ vs. $k_y$ for the MWSM$||$ child created from the above two parents for $k_z=0$ and $k_z=\frac{\pi}{2}$ respectively. (g) shows the slab spectra $E$ vs. $k_z$ at $k_y=0$ for the same MWSM$||$ child system.}
    \label{MWSMpllfinitex05m05}
\end{figure}

\begin{figure}
    \hspace*{-1.0cm}
    \includegraphics[scale=0.6]{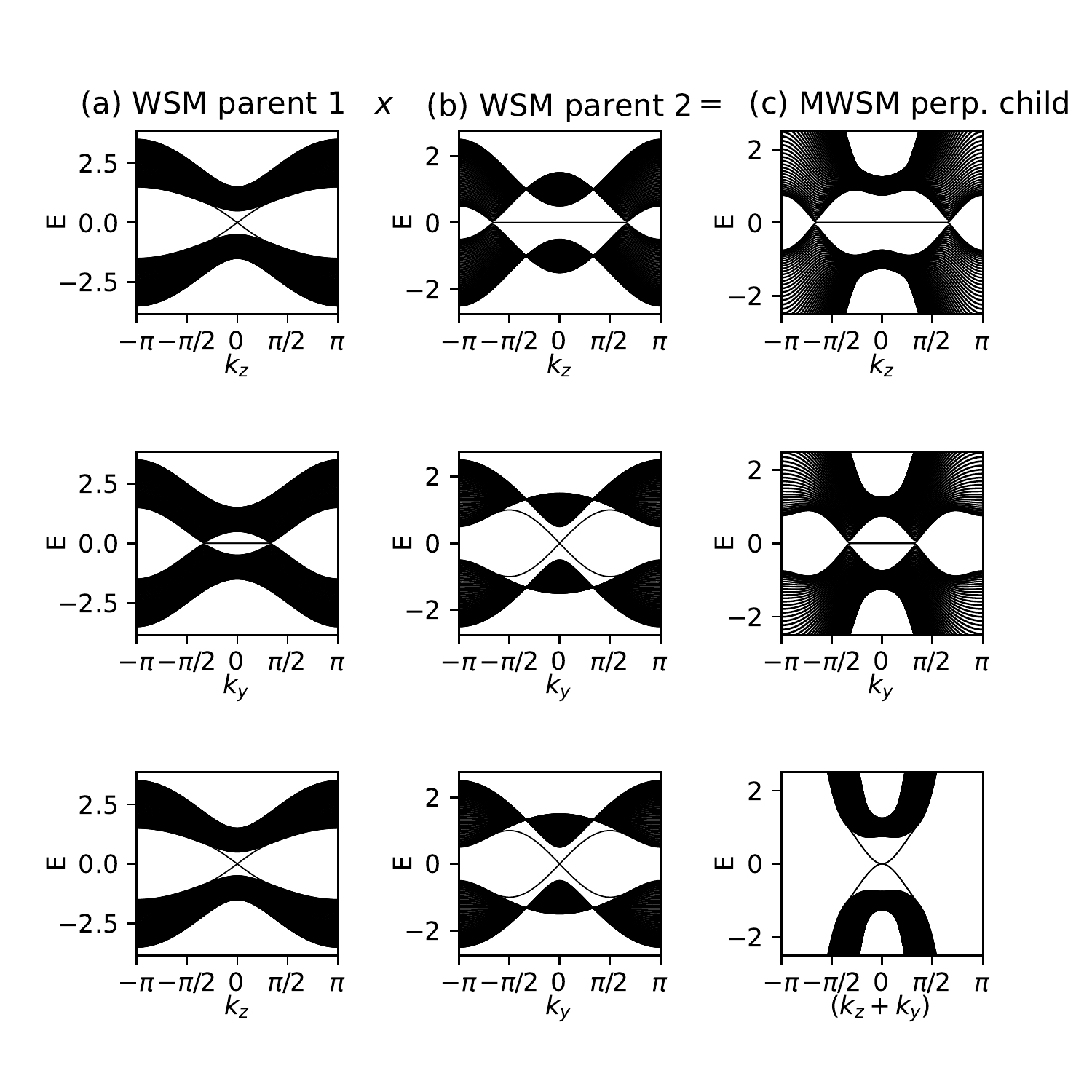}
    \caption{Finite slab spectra (in x direction, $N_x=80$) with the constituent parent Hamiltonians - WSM parent Hamiltonian 1 with $\gamma_1=0.5$ and Weyl nodes along the $k_y$-direction is shown along column (a), WSM parent Hamiltonian 2 with $\gamma=-0.5$ with Weyl nodes along the $k_z$-direction along column (b) and the MWSM perpendicular child Hamiltonian along column (c). It is apparent how the surface spectra along $k_z$(for $k_y=0$) and $k_y$(for $k_z=0$) combine multiplicatively to create the surface spectra for the MWSM perpendicular system. The lowest diagram along column (c) especially shows the spectra along the diagonal $k_z+k_y$ direction where the component spectra $\sin(k_z)$ and $\sin(k_y)$ have combined to produce $\sin(k_z)\sin(k_y)$ as the leading term.}
    \label{MWSMperpfinitex05m05allk}
\end{figure}

One might expect that the tensor product structure of the multiplicative phases is visible in the surface spectrum of the MWSM. Numerical simulations show that this is the case. For the parent WSMs, the surface spectra is given as, $E(k_y)\sim \sin(k_y)$(Fig.~\ref{MWSMpllfinitex05m05}(a) and (b) and Fig. \ref{MWSMperpfinitex05m05allk} 2nd row) for nodes along the z-axis and open boundaries along the x-direction and $E(k_z)\sim \sin(k_z)$(Fig.~\ref{MWSMperpfinitex05m05allk} 1st row) for nodes along the y-axis and open boundaries along the x-direction. Indeed, corresponding surface spectra of child Hamiltonians depend on these surface spectra in a multiplicative way. Numerical simulation from  Fig.~\ref{MWSMpllfinitex05m05} (c) shows that, for MWSM$||$, the slab spectra disperses as $E(k_y)\sim \sin^2(k_y)$ for two parents each with surface spectrum $E(k_y)\sim \sin(k_y)$. In contrast, Fig.~\ref{MWSMperpfinitex05m05allk} (c) shows that the surface spectrum instead disperses as $E(k_y,k_z)\sim \sin(k_y)\sin(k_z)$ for MWSM$\perp$ when one parent has the former surface spectrum and the other has the latter. We also show, for the case of each parent surface spectrum along $k_z$, which exhibits flat bands between the two Weyl nodes (Fig.~\ref{MWSMpllfinitex05m05}(a) and (b)) corresponds to flat bands between all four gapless points in the MWSM parallel system (Fig.~\ref{MWSMpllfinitex05m05}(c)). However, fitting $\sin^2(k_y)$ curves to each of the parallel and perpendicular MWSM spectra reveals that, except in special cases when $\gamma_1=\gamma_2$ where the fit is exact, the slab spectra does not disperse as $\sin^2(k_y)$ and instead exhibits $k_z$-dependence. One can check this by comparing $E$ vs. $k_y$ slab spectra in the range $-\text{min}(k_{0,1},k_{0,2})<k_z<\text{min}(k_{0,1},k_{0,2})$ and $\text{min}(k_{0,1},k_{0,2})<k_x<\text{max}(k_{0,1},k_{0,2})$. The spectra appears linear near zero in the latter case.

\subsubsection{Stability of surface states of MWSM}

For the MWSM$||$ system, the low-energy expansion about a node is reminiscent of a Dirac node, and it is therefore possible to break apart the four-fold degeneracy at each of the nodes by introducing an external magnetic field. We introduce minimal coupling, $k_y\rightarrow k_y-eBx$ for the MWSM$||$ to simulate the effect of applied magnetic field on the spectral density of the Fermi arc surface states. We observe that the Fermi arcs split but are not destroyed by the applied field as in the case of the DSM.\\

\subsubsection{Fermi Arcs for the MWSM as a stack of MCIs:}
WSMs can be interpreted as a set of Chern insulators(CIs), each defined in a 2d submanifold of the 3d BZ of the WSM (e.g., each $k_x$-$k_y$ plane) for a given value of $k_z$, yielding a stack of CIs in the $k_z$-direction. The Weyl nodes then correspond to topological phase transitions---corresponding to gap-closings---in the stack between intervals in $k_z$ with topologically-distinct CIs. Specifically, we use the SCZ model \cite{qi2006topological}, $\mc{H}_{CI}=B(2+M-\cos k_x -\cos k_y)\sigma^z+\sin k_x\sigma^x+\sin k_y\sigma^y$ in particle-hole space. In the WSM, the mass term is given as $M=\gamma-\cos k_z$. Here, for the range, $-1<\gamma <1$, $k_z\in [-\cos^{-1}\gamma, \cos^{-1}\gamma]$. The Fermi arcs we observe in the 2d BZ defined in the $k_y-k_z$ for open boundary conditions in the x-direction are projections of the chiral edge states of the slices of the corresponding CIs in the stack.\\

\begin{center}
\begin{figure*}
\centering
\includegraphics[scale=0.6]{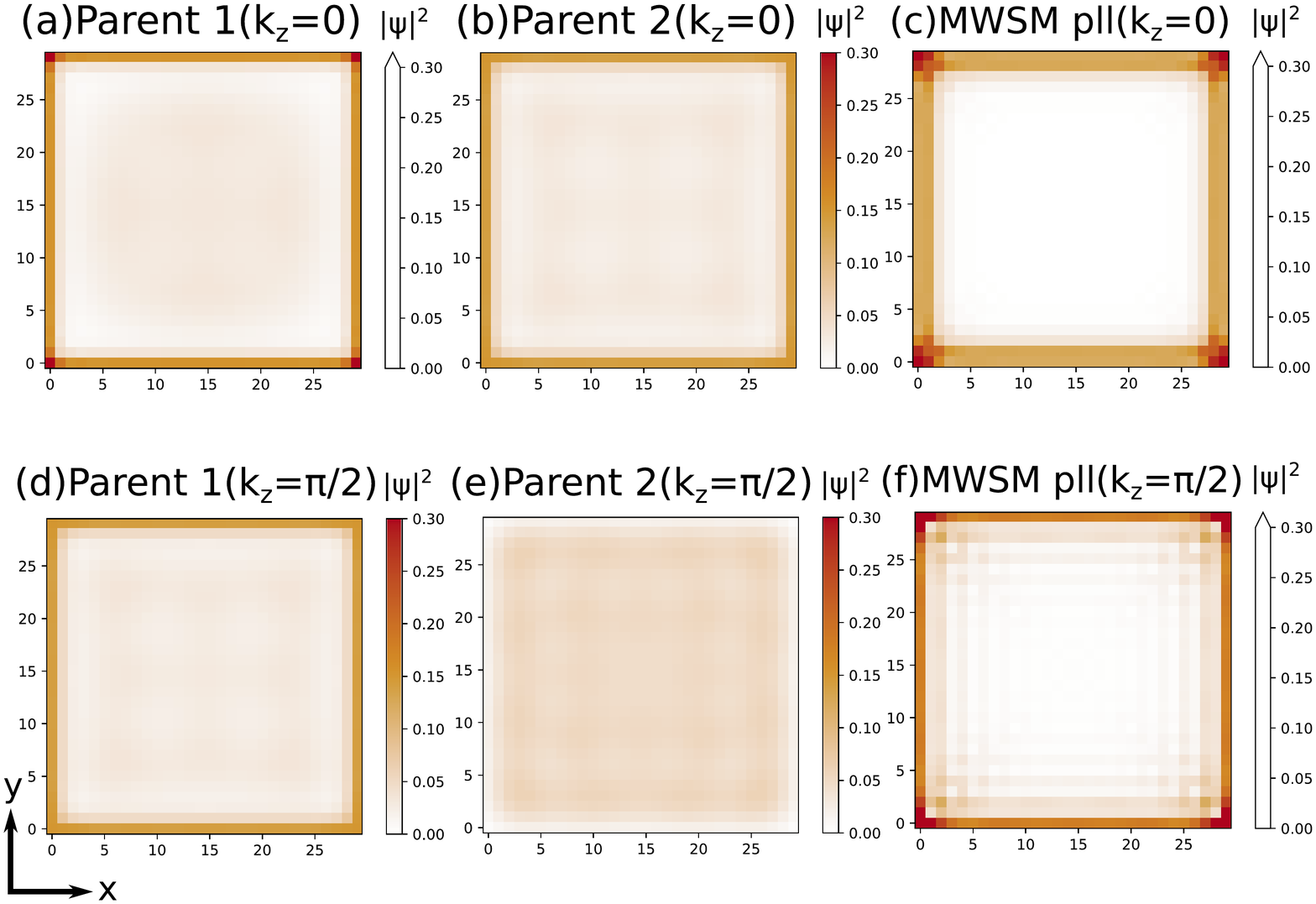}
\caption{Probability densities of superposition of 40 edge state eigenvectors in a $30\times 30$($L_x\times L_y$) square lattice at $k_z=0$ and $k_z=\frac{\pi}{2}$ for (a, d) Parent WSM 1 ($\gamma_1=-0.5$), (b, e) Parent WSM 2($\gamma_2=0.5$) and (c, f) MWSM $||$ child ($\gamma_1=-0.5$ and $\gamma_2=0.5$) respectively. At $k_z=0$, both the parent systems are topological as seen from a visible edge state which results in localization at both the edge and second last edge sites in the MWSM $||$ child system. When $k_z=\frac{\pi}{2}$, the parent 1 is still topological but the parent 2 is trivial as seen from the absence of edge states which results in localization only at the edge sites of the MWSM $||$ child system.}
\label{MWSMplledgepdf}
\end{figure*}
\end{center}

The multiplicative counterpart of a Chern insulator was introduced recently by Cook and Moore~\cite{cook2022mult} as Multiplicative Chern Insulators(MCIs). Here, one must notice that the MCI has two mass terms derived from each of the parent systems, one from each of the parent systems. Hence, there exists more than one way to stack the MCIs in the $k_z$ direction. Either parent mass term can be $k_z$-dependent, for instance, or both can be. Here, we have attached the momentum dependence to both the mass terms, so that the difference in parent mass parameters remains constant. We then characterize the multiplicative Fermi arc states by opening boundary conditions in the x- and y-directions, and plotting the probability density for the sum of 40 eigenstates nearest in energy to zero in Fig. \ref{MWSMplledgepdf} for $k_z=0$ (a 2D submanifold of the BZ realizing an MCI) and $k_z=\frac{\pi}{2}$ (a 2D submanifold of the BZ that is topologically trivial). For the former case shown in Fig.~\ref{MWSMplledgepdf}(a) and (b), the probability density in the corresponding child is localized at sites at the boundary, but also at the sites adjacent to these sites. For the latter case, parent 1 has edge states and parent 2 does not as shown in Fig.~\ref{MWSMplledgepdf}(d) and (e). The resultant child probability density shows low-energy states localize only at the boundary sites as shown in Fig.~\ref{MWSMplledgepdf}(f). This localization behavior is similar to that of the multiplicative Kitaev chain presented in a second work by the present authors, where, if each parent is topological, edge states are localized at lattice sites right at the edge, but also at sites adjacent to these sites. We expect such localization to protect the edge states from backscattering to some extent, which we will explore in future work.

\subsubsection{Boundary states disconnected from bulk states---}
The MCI introduced by Cook and Moore~\cite{cook2022mult} can exhibit topologically robust yet floating edge states, which are separated from the bulk by a finite energy gap. MTSMs constructed from MCIs can inherit this exotic boundary state connectivity, displaying boundary states disconnected from the bulk band structure.

To realize such a MWSM, we first note when edge states are disconnected from bulk states for the case of the MCI:
\begin{subequations}
\begin{equation}
\begin{split}
\mc{H}_{CI,p1}(\boldsymbol{k}) =& B_1(2+M_1-\cos k_x-\cos k_y)\tau^z\\
&+\sin k_x\tau^x+\sin k_y\tau^y,
\end{split}
\label{Hci1par}
\end{equation}
\begin{equation}
\begin{split}
\mc{H}_{CI,p2}(\boldsymbol{k}) =& B_2(2+M_2-\cos k_x-\cos k_y)\sigma^z\\
&+\sin k_x\sigma^x+\sin k_y\sigma^y,
\end{split}
\label{Hci2par}
\end{equation}
\begin{equation}
\begin{split}
\mc{H}_{MCI,c}(\boldsymbol{k}) =& [B_1(2+M_1-\cos k_x-\cos k_y)\tau^z\\
&+\sin k_x\tau^x+\sin k_y\tau^y]\\
&\otimes [-B_2(2+M_2-\cos k_x-\cos k_y)\sigma^z\\
&-\sin k_x\sigma^x+\sin k_y\sigma^y],
\end{split}
\label{Hmcipll}
\end{equation}
\end{subequations}
the range of parameters over which this is possible is $M_1\in [-4,-2]$ and $M_2\in [-2,0]$ which corresponds to Chern numbers $C=+1$ and $C=-1$ respectively. We therefore construct a MWSM for which the Weyl nodes of one parent WSM are separated in k-space by a stack of Chern insulators, each with total Chern number $C=+1$, and the Weyl nodes of the other parent are separated by a stack of Chern insulators, each with total Chern number $C=-1$. Comparing Eqn.~\eqref{Hws1par} with \eqref{Hci1par} and Eqn.~\eqref{Hws2par} with \eqref{Hci2par}, it is clear that, for each Chern insulator in the stack, the following mapping holds, $M_i = \gamma_i-\cos k_z$, $i\in\{1,2\}$, and $i$ labeling the parent. From this mapping, it is not possible to have $M_2\in (-4,-2)$ while $\gamma_i\in (-1,1)$, $i\in\{1,2\}$. We therefore generalize the mapping to the following form, $M_i=\gamma_i-r_i\cos k_z$, $i\in\{1,2\}$, so that the parents and the child Hamiltonian for the MWSM parallel are,
\begin{subequations}
\begin{equation}
\begin{split}
\mc{H}_{p1}(\boldsymbol{k})=& t_{11}\sin k_x\tau^x+t_{21}\sin k_y\tau^y\\
& +t_{31}(2+\gamma_1-\cos k_x-\cos k_y-r_1\cos k_z)\tau^z,\\
\end{split}
\label{Hws1par}
\end{equation}
\begin{equation}
\begin{split}
\mc{H}_{p2}(\boldsymbol{k})=&t_{12}\sin k_x\sigma^x+t_{22}\sin k_y\sigma^y\\
&+t_{32}(2+\gamma_2-\cos k_x-\cos k_y-r_2\cos k_z)\sigma^z,
\end{split}
\label{Hws2par}
\end{equation}
\begin{equation}
\begin{split}
\mc{H}_c(\boldsymbol{k})=& [t_{11}\sin k_x\tau^x+t_{21}\sin k_y\tau^y\\
& +t_{31}(2+\gamma_1-\cos k_x-\cos k_y-r_1\cos k_z)\tau^z]\\
& \otimes [-t_{12}\sin k_x\sigma^x+t_{22}\sin k_y\sigma^y\\
&-t_{32}(2+\gamma_2-\cos k_x-\cos k_y-r_2\cos k_z)\sigma^z].
\end{split}
\label{Hmwspll}
\end{equation}
\end{subequations}

\begin{figure}[h!]
\centering
\hspace{-1cm}
\includegraphics[width=0.53\textwidth]{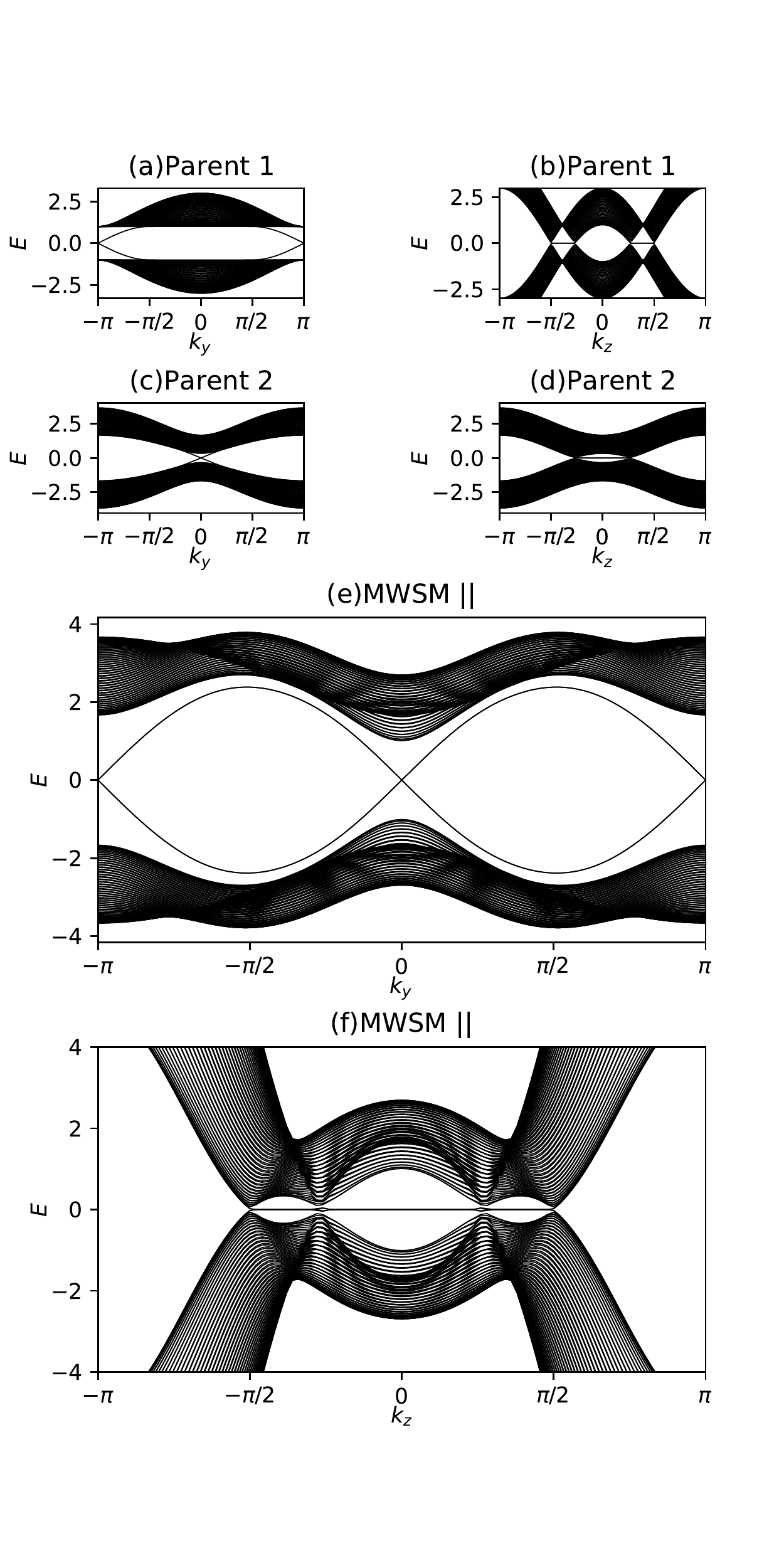}
\caption{Slab spectra along $k_y$ (subfigure a)  and $k_z$ (subfigure b) for WSM parent 1 with $\gamma_1=0$, $r_1=3$, and slab spectra along  $k_y$ (subfigure c) and $k_z$ (subfigure d) for WSM parent 2 with $\gamma_2=2/3$, $r_2=1$, respectively. Corresponding slab spectra for the MWSM$||$ with $t_{11}=t_{12}=1$, $t_{21}=t_{22}=1$, $t_{31}=t_{32}=1$ along (e) $k_y$  and (f) $k_z$, respectively, with edges separate from the bulk slab spectra along $k_y$.}
\label{MWSMplledgeminusbulkg0r3g067r1}
\end{figure}

To construct one parent with Chern number of this stack non-trivial and opposite in sign to the Chern number of the stack in the other parent, we first introduce some terminology. We refer to the region between Weyl nodes including $k_z=0$ as regular Weyl region (RWR) and the region including $k_z=\pm\pi$ as the irregular Weyl region (IWR). The existence of Weyl nodes requires $|r_{1,2}|\geq 1$ for $|\gamma_{1,2}|<1$. It is then possible to realize a RWR with negative Chern number by varying $r_{1,2}$, so that $\gamma_{1,2}-r_{1,2}\cos k_z\in (-4,-2)$. These RWRs---one of each parent system---must then occur over the same interval in $k_z$, however, to realize topological floating surface states. We set $\gamma_2=0$ and $r_2=3$, which means we have $C=-1$ for the range $[-\cos^{-1}(\frac{2}{3}),\cos^{-1}(\frac{2}{3})]$ when $M_2=\gamma_2-r_2\cos k_z\in [-3,-2]$. Then we must have $\gamma_1=\cos\frac{\pi}{3}=0.5$ and $r_1=1$ so that in the region $k_z\in [-\cos^{-1}(\frac{2}{3}),\cos^{-1}(\frac{2}{3})]$, we have the same kind of MCI with edge states gapped from the bulk as described in Cook and Moore\cite{cook2022mult}. These results are shown in Fig.~\ref{MWSMplledgeminusbulkg0r3g067r1}.\\


The MWSM$\perp$ case of topologically robust yet floating Fermi arc surface states is constructed similarly, and we defer thorough investigation of this case to later work.


\section{Effect of Magnetic field on MWSM and Chiral anomaly}
We now investigate response signatures of MTSMs. As we consider MWSMs here, which may be constructed from WSM parent systems, we focus in particular on the question of whether there is a multiplicative generalization of the chiral anomaly, one of the most important signatures of Weyl semimetals: application of non-orthogonal electric and magnetic fields can pump electrons between Weyl nodes of opposite chirality~\cite{jia_weyl_2016}. More specifically, applying an external magnetic field parallel to the axis along which Weyl nodes are separated in k-space yields a single chiral Landau level near each of the Weyl nodes. In Weyl semimetals, this suppresses backscattering of electrons with opposite chirality, manifesting as a negative magnetoresistance (MR). Weyl semimetals therefore serve as condensed matter platforms for study of the chiral anomaly, also known as Adler-Bell-Jackiw anomaly, associated with the Standard Model of particle physics~\cite{yanreview2017}. When the external magnetic field is instead oriented perpendicular to the k-space axis along which Weyl nodes are separated, semi-classical calculations indicate the presence of quantum oscillations in the density of states~\cite{potter_quantum_2014}, observable in magnetization, magnetic torque, and MR measurements~\cite{yanreview2017}.

To study the effects of external fields on the MWSM, we first derive the Landau level structure for the the Weyl semimetal in the cases of external magnetic fields applied parallel and perpendicular to the Weyl node axis. We can then draw parallels between these results and their generalizations in the case of the MWSM.

\subsection{Chiral anomaly in WSM}
\begin{figure}[h!]
    \includegraphics[width=0.45\textwidth]{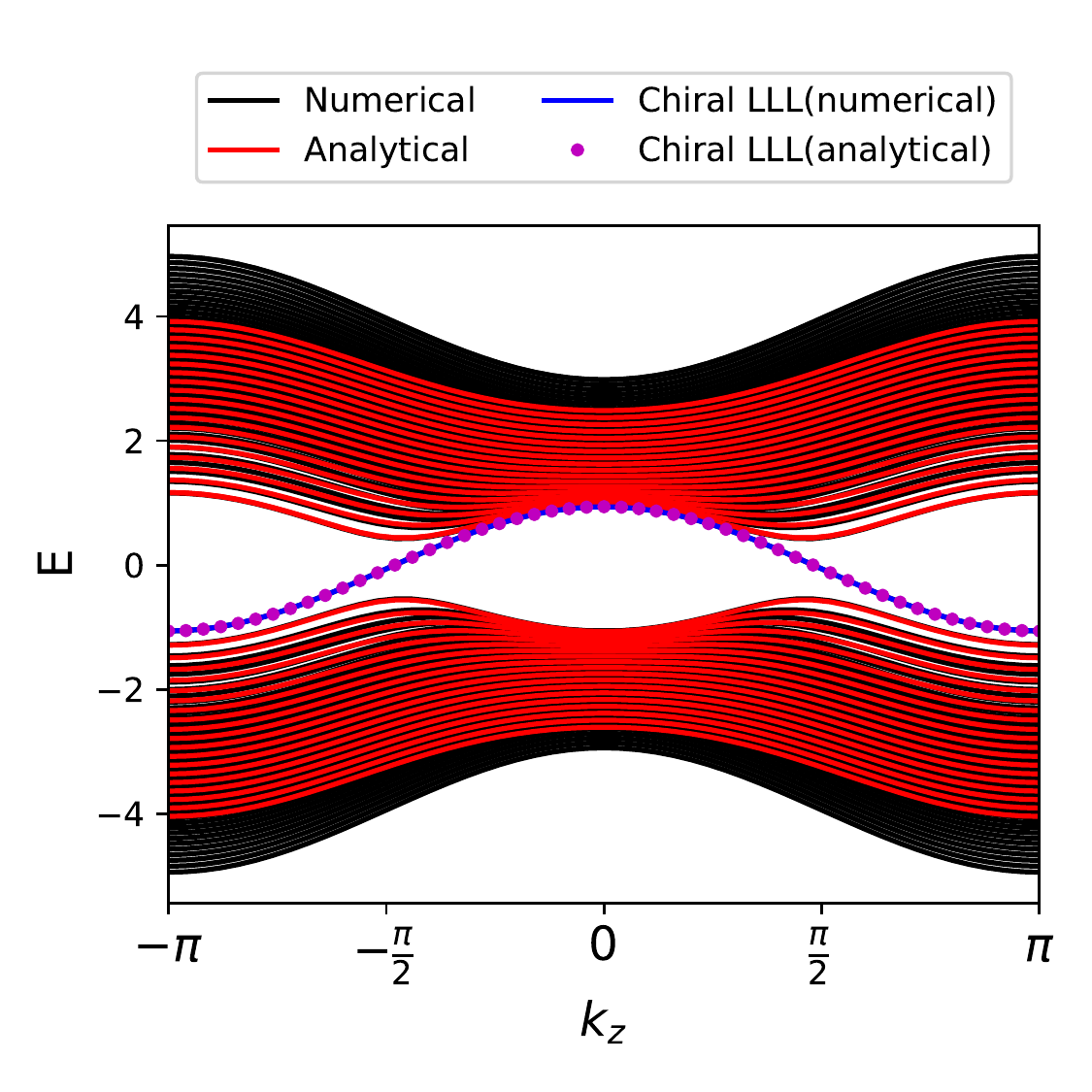}
    \includegraphics[width=0.45\textwidth,trim={0 0 0.6cm 2.5cm},clip]{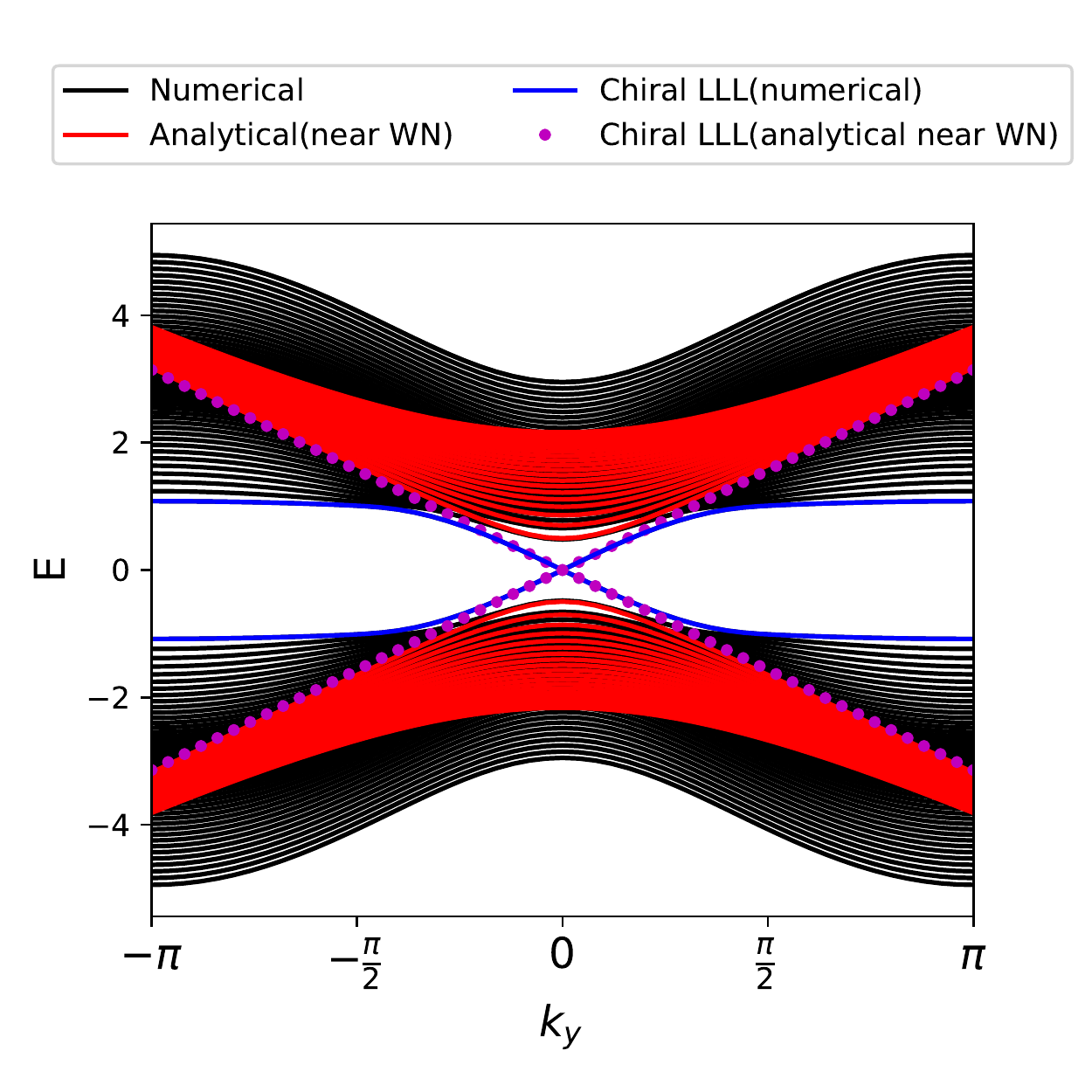}
    \caption{Landau Levels for the two-band Weyl Semimetal calculated analytically from Eqn. \ref{WSMJC} and numerically, with $t=1=t^\prime$, $\gamma=0$ and $\mathbf{B}=\frac{2\pi}{51}\hat{\mathbf{z}}$(upper) and $\mathbf{B}=\frac{2\pi}{51}\hat{\mathbf{y}}$(lower). The (black) bands indicate the numerically calculated Landau levels and the (red) bands for the analytically calculated Landau levels for $n=1,2,...,19$. The (blue) band and the dotted (magenta) band is the Lowest Landau Level(LLL) calculated numerically and analytically, and is responsible for the Chiral anomaly in the upper figure and Weyl orbits in the lower figure.}
    \label{WSMLandauLevels}
\end{figure}

To study the chiral anomaly in a WSM, we consider a particular Bloch Hamiltonian $H_{WSM}(\boldsymbol{k})$ characterizing a Weyl semimetal phase and its expansion around the $k_z$-axis, i.e. $\boldsymbol{k}\rightarrow (0,0,k_z)$ (up to 2nd order in $k_x$ and $k_y$),
\begin{equation}
\begin{split}
H_{WSM}(\boldsymbol{k}) =& t(2+\gamma-\cos k_x-\cos k_y-\cos k_z)\sigma^z\\
&+t^\prime\sin k_y\sigma^y+t^\prime\sin k_x\sigma^x,\\
\approx & t(Q+\frac{1}{2}(k_x^2+k_y^2))\sigma^z+t^\prime k_y\sigma^y+t^\prime k_x\sigma^x,
\end{split}
\label{WSMminimal}
\end{equation}
where $Q=\gamma-\cos k_z$. Applying the magnetic field, $\mathbf{B}=B\hat{\mathbf{z}}$ along the Weyl node axis, Peierls substitution changes the momenta in the following way, $k_x\rightarrow k_x^\prime = k_x$, $k_y\rightarrow k_y^\prime = k_y+eBx$, and $k_z\rightarrow k_z^\prime = k_z$. The position-momentum commutator, implies, $[k_y^\prime,k_x^\prime] = ieB$, so that, it is possible to define bosonic ladder operators,
\begin{equation}
\begin{split}
a = \frac{k_x^\prime - ik_y^\prime}{\sqrt{2eB}};\quad a^\dagger = \frac{k_x^\prime + ik_y^\prime}{\sqrt{2eB}};\quad [a,a^\dagger] = 1.
\end{split}
\label{ktoboson}
\end{equation}
Applying Eqn.\ref{ktoboson}, after substituting $\boldsymbol{k}\rightarrow \boldsymbol{k}^\prime$, we get the following system which looks similar to the polariton conserving Jaynes-Cummings Hamiltonian,
\begin{equation}
H_{WSM}(\boldsymbol{k}^\prime)\approx t(Q+eB(a^\dagger a+\frac{1}{2}))\sigma^z+t^\prime\sqrt{2eB} (a\sigma^++a^\dagger\sigma^-),
\label{WSMJC}
\end{equation}
where $\sigma^\pm = \frac{1}{2}(\sigma^x\pm i\sigma^y)$ are the spin ladder operators in the basis $\{\ket{+},\ket{-}\}$ of $\sigma^z$ ($\sigma^z\ket{\pm}=\pm\ket{\pm}$). The ground state from the above Hamiltonian is given by the eigenvector, $\ket{\psi_{LLL}} = \ket{0;-}$ (states denoted as $\ket{n;s}$ where n is the bosonic number and $s$ is the spin direction), which leads to the lowest Landau level energy,
\begin{equation}
E_{LLL} = -t(Q+\frac{1}{2}eB).
\end{equation}
Near each of the Weyl nodes, it is easy to observe that $\ket{\psi_{LLL}}$ is chiral as shown in Fig. \ref{WSMLandauLevels}. The other Landau levels can be derived by restricting to the two dimensional disjoint spaces, $\{\ket{n,-},\ket{n-1,+}\}$, parametrized by the bosonic number, $n$ so that in each such basis, the Hamiltonian is,
\begin{equation}
H(k_z,n) = -\frac{teB}{2}\sigma^0-t(Q+eBn)\sigma^z+t^\prime\sqrt{2eBn}\sigma^x.
\label{WSMLL2band}
\end{equation}
The energy for the other Landau levels parametrized by $n=1,2,...$ is given by the eigenvalues of Eqn. \ref{WSMLL2band},
\begin{equation}
E_{nLL} = -\frac{teB}{2}\pm \sqrt{t^2(Q+eBn)^2+2{t^\prime}^2eBn}.
\end{equation}
We have illustrated the analytically calculated Landau levels in Fig. \ref{WSMLandauLevels} and compared them to numerical calculations of Landau levels. The numerical computation involves plotting the bands for the Peierls substituted Weyl semimetal with periodic boundary conditions, say in the x-direction, and subjected to magnetic field in integer multiples of $\frac{2\pi}{L}$, where L is the size of the lattice in the x-direction \cite{}. We observe that the chiral Landau level from both analytical and numerical methods overlap, with an approximate overlap of the other Landau levels since we only considered till second order in $k_x$ and $k_y$.

Next we consider the case when the magnetic field is directed perpendicular to the Weyl node axis, say $\mathbf{B}=B\hat{\mathbf{y}}$. Expanding the first line of Eqn. \ref{WSMminimal} around the Weyl node, $\boldsymbol{k}=(0,0,k_0=\cos^{-1}\gamma)$ of positive chirality, and setting $t=t^\prime = 1$, we get,
\begin{equation}
\begin{split}
H_{WSM}(\boldsymbol{k})\approx & \sin k_0(k_z-k_0)\sigma^z+k_y\sigma^y+k_x\sigma^x,\\
\implies H^\prime_{WSM}(\boldsymbol{k}) \approx & -k_y\sigma^z+k_x\sigma^x+\sin k_0(k_z-k_0)\sigma^y,
\end{split}
\label{WSMBy}
\end{equation}
where in the second line we have rotated the Hamiltonian to a new basis via, $\sigma^x\rightarrow\sigma^z$ and $\sigma^x\rightarrow -\sigma^x$. In the presence of mentioned magnetic field perpendicular to the Weyl node axis, the Peierls substitution is applied as $k_x\rightarrow k_x^\prime=k_x$, $k_y\rightarrow k_y^\prime = k_y$ and $k_z\rightarrow k_z^\prime = k_z-eBx$. The commutation relation, $[k_x,\sin k_0(k_z-k_0-eB_x)]=ieB\sin k_0$, then constructs the bosonic ladder operators,
\begin{equation}
b = \frac{k_z-k_0-eBx-ik_x}{\sqrt{2eB\sin k_0}}; \quad b^\dagger = \frac{k_z-k_0-eBx+ik_x}{\sqrt{2eB\sin k_0}}.
\end{equation}
The system in  Eqn. \ref{WSMBy} then changes to,
\begin{equation}
H_{WSM}(\boldsymbol{k}^\prime)\approx -k_y\sigma^z+\sqrt{2eB\sin k_0}(b\sigma^++b^\dagger\sigma^-).
\label{WSMByJC}
\end{equation}
Similar to the previous case, it is possible to resolve the Hamiltonian into the subspaces spanned by $\{\ket{n,-},\ket{n-1,+}\}$, where $n$ is the eigenvalue of the number operator, $b^\dagger b$. We get two chiral lowest Landau levels with energies, $E = \pm k_y$ in the bulk, which are responsible for the chiral anomaly~\cite{yanreview2017}.

\subsection{Chiral anomaly in the MWSM}
We now study the response of the MWSM to external fields for comparison with the signatures of the chiral anomaly in the WSM reviewed in the previous section. We treat the MWSM parallel and perpendicular cases separately, given the expected sensitivity of the response to orientation of the axes of node separation relative to the orientation of the external fields.

\subsubsection{Landau levels in the MWSM parallel system:}
\begin{figure}[h!]
    \centering
    \includegraphics[scale=0.5]{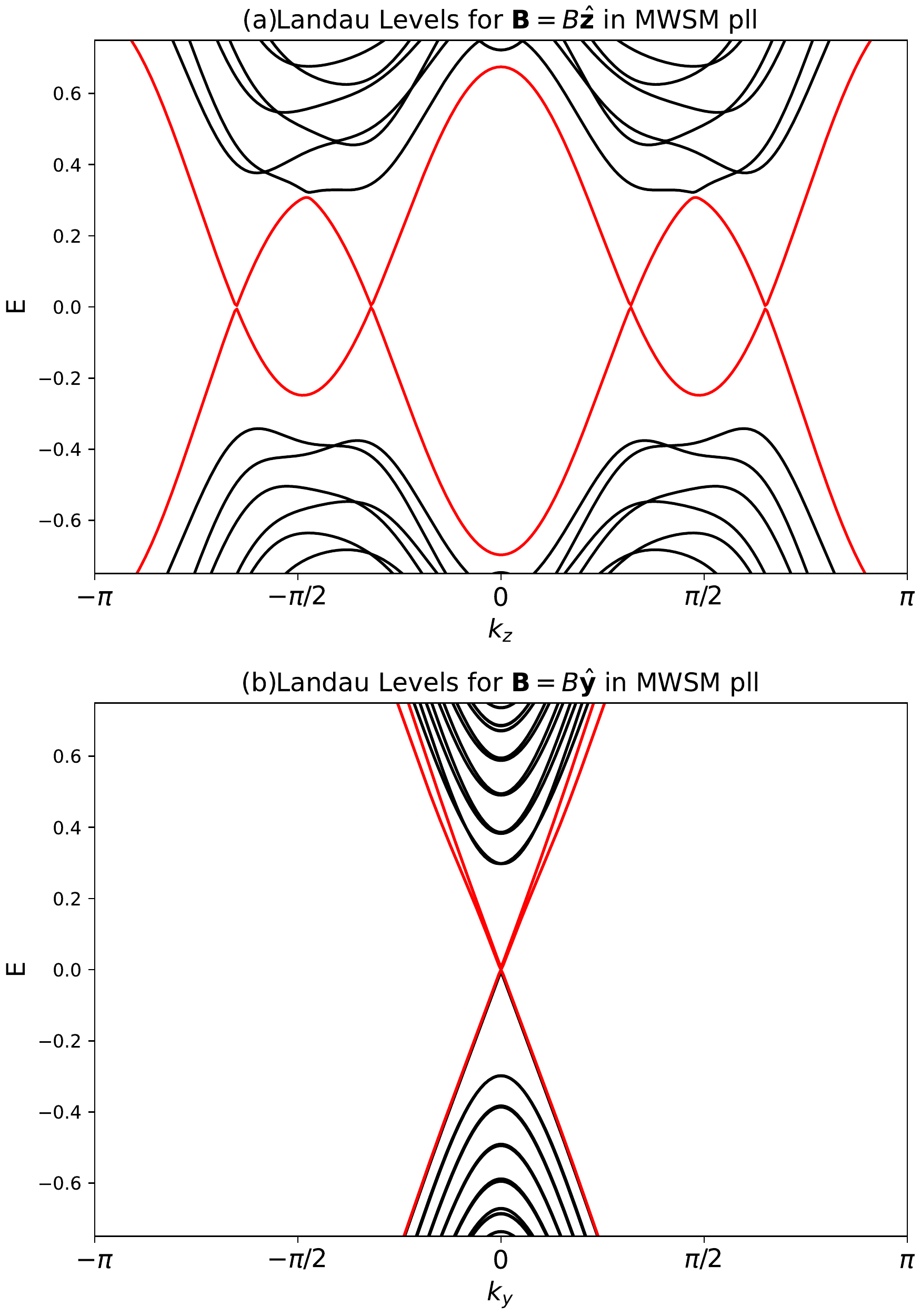}
    \caption{The Landau Levels for the MWSM parallel Hamiltonian with $\gamma_1=-0.5$, $\gamma_2=0.5$, $t_1=t_1^\prime =t_2=t_2^\prime = 1$ and $B=\frac{2\pi}{80}$. (a) and (b) show the Landau levels for the magnetic field along the Weyl axis and perpendicular to the Weyl axis (at Weyl node $(0,0,\frac{\pi}{3})$ respectively. The (red) bands refer to the lowest Landau levels and the (black) bands form the bulk Landau levels.}
    \label{MWSM_Landau_Levels}
\end{figure}

In Sec. \ref{MWSMpllintro} we have derived the Dirac Hamiltonian for the MWSM$||$ in the vicinity of each of its two nodes, $(0,0,k_{01})$ and $(0,0,k_{02})$ derived respectively from each of its two parents.
$$
\begin{aligned}
H^c_{||,1}(\boldsymbol{k}) =& (t_{1}^\prime k_x\tau^x+t_{1}^\prime k_y\tau^y\\
&+t_{1}\sin k_{01}\bar{k}_{z,1}\tau^z)t_{2}(\gamma_1-\gamma_2)\sigma^z,\\
H^c_{||,2}(\boldsymbol{k}) =& t_{1}(\gamma_1-\gamma_2)\tau^z(-t_{2}^\prime k_x\sigma^x+t_{2}^\prime k_y\sigma^y\\
&-t_{2}\sin k_{02}\bar{k}_{z,2}\sigma^z)
\end{aligned}
$$
In this section, we will only consider cases where $\gamma_1\neq \gamma_2$. To investigate the response to external fields for the MWSM$||$, we consider the effect of magnetic field along the Weyl node axis, i.e., $\mathbf{B}=B\hat{\mathbf{z}}$. We use the exact Peierls substitution in Eqn. \ref{ktoboson}, so that the two expressions above transform as follows,
\begin{equation}
\begin{split}
H^c_{||,1}(\boldsymbol{k}^\prime) =& t_2(\gamma_1-\gamma_2)(t_1\sin k_{01}\bar{k}_{z,1}\tau^z\\
&+t_1^\prime\sqrt{2eB}(a\tau^++a^\dagger\tau^-))\sigma^z,\\
H^c_{||,2}(\boldsymbol{k}^\prime) =& t_1(\gamma_1-\gamma_2)\tau^z(-t_2\sin k_{01}\bar{k}_{z,2}\sigma^z\\
&-t_2^\prime\sqrt{2eB}(a\sigma^-+a^\dagger\sigma^+)).
\end{split}
\end{equation}
Here $\tau^{\pm}=\frac{1}{2}(\tau^x\pm i\tau^y)$ and $\sigma^\pm = \frac{1}{2}(\sigma^x\pm i\sigma^y)$ are the pseudo-spin ladder operators in the $\tau$ and $\sigma$ spaces. The lowest Landau levels from the above two expressions are given below,
\begin{equation}
\begin{split}
H^c_{||,1}\rightarrow & E_{1,LLL} = \pm(\gamma_1-\gamma_2)t_1t_2\sin k_{01}\bar{k}_{z,1},\\
&\ket{\psi_{1,LLL}} = \ket{0;-,\pm},\\
H^c_{||,2}\rightarrow & E_{2,LLL} = \mp(\gamma_1-\gamma_2)t_2t_2\sin k_{02}\bar{k}_{z,2},\\
&\ket{\psi_{2,LLL}} = \ket{0;\pm,+}.
\end{split}
\end{equation}
One may notice that the eigenvector $\ket{0;-,+}$ occurs in the vicinity of each node. Therefore, we calculate its energy eigenvalue if one expands the MWSM parallel system in the vicinity of the $k_z$ axis. The details of the calculation can be found in the Supplementary Materials S3. We find the energy is given as,
\begin{equation}
E_{\ket{0;-,+}} = (Q_1Q_2+\frac{1}{2}eB(Q_1+Q_2)).
\end{equation}
We show that this expression is consistent with the numerically calculated Landau levels in Fig.~\ref{MWSM_Landau_Levels}. The other chiral Landau level consistent with the other two eigenvectors, $\ket{0;-,-}$ and $\ket{0;+,+}$ near their respective Weyl nodes appears distinct from $\ket{0;-,+}$ away from the Weyl nodes.

In Fig.~\ref{MWSM_Landau_Levels}, for certain values of $\gamma_1$ and $\gamma_2$, it appears, at first glance, as if there are two separate, chiral Landau levels corresponding to $\ket{0;-,-}$ and $\ket{1;,-,-}$ respectively. All four Weyl nodes are connected by each of these LLLs, however, and the two LLLs in combination furthermore account for each chirality at each node. Although this is reminiscent of the Dirac semimetal, there is potentially a distinction in character between the chiralities at each node. If each parent corresponds to a particular degree of freedom, for instance, and these dofs are physically distinct from one another in some sense, such as one parent corresponding to a two-fold valley dof, and the other corresponding to a two-fold layer dof, the chiral anomalies are inequivalent and do not compensate one another as they would for a Dirac semimetal.

The two apparently 'separated' LLLs seem to only scatter between the Weyl nodes derived from their respective parents, i.e. intra-parent scattering. Upon closer inspection, however, we see the intersection point between two apparently separated Landau levels is actually a very small gap. We have verified in Supplementary Sec.~S3, that the gap is finite in analytical calculations performed to second order in momenta. The gap is an emergent feature of the multiplicative chiral anomaly, with the single LLL reducing to $\ket{0;-,-}$ and $\ket{0;+,+}$ at nodes associated with a particular parent. We therefore interpret the multiplicative chiral anomaly as exhibiting parent-graded features as well as emergent features not associated with either individual parent. This is reminiscent of the topologically robust floating bands of the multiplicative Chern insulator~\cite{cook2022mult}.

\subsubsection{Landau levels in the MWSM perpendicular system:}
\begin{center}
\begin{figure}[htb!]
\centering
\includegraphics[scale=0.45]{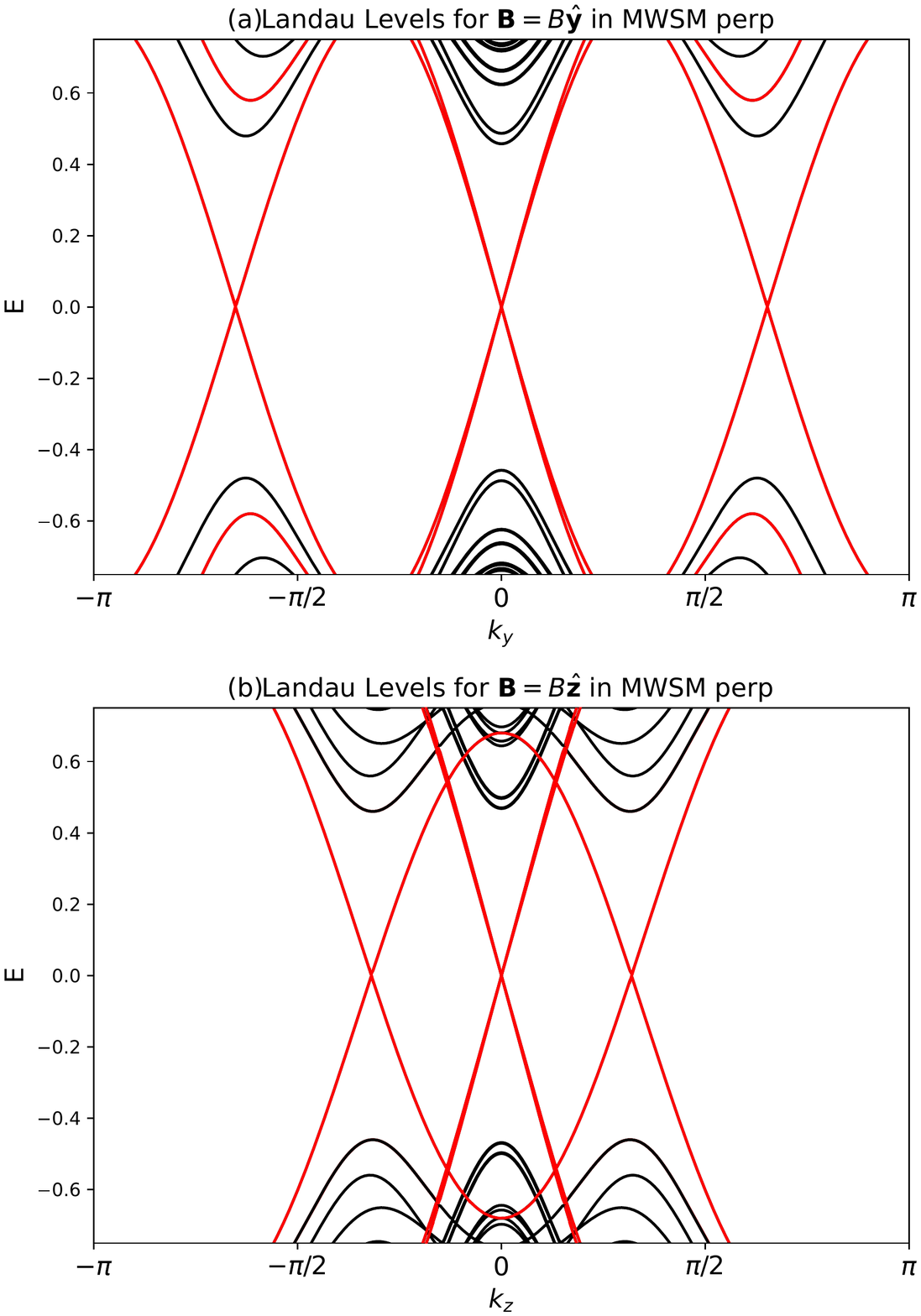}
\caption{Landau levels for the MWSM perpendicular system with $\gamma_1=-0.5$ and $\gamma_2=0.5$ representing separation of Weyl nodes along the $k_y$ and $k_z$ direction respectively. We show two cases, (a) when magnetic field is along the y-direction and (b) when magnetic field is along the z-direction. Red lines indicate the chiral Landau levels. since the magnetic field is paralle to one Weyl node separation and perpendicular to another Weyl node separation, the above behaviour is expected.}
\label{MWSMperpLandaulevels}
\end{figure}
\end{center}

In Sec.~\ref{MWSMperpbulk}, we had shown the linear expansion of the MWSM$\perp$ Bloch Hamiltonian near each of the nodes corresponding to one parent with Weyl nodes separated along the $k_y$ axis and the other parent with Weyl nodes separated along the $k_z$ axis in Eqn. \ref{MWSMperpliny} and \ref{MWSMperplinz}. Without loss of generality, we consider, $t_{31}=t_{32}=t_{21}=t_{22}=1=t_{11}=t_{12}$. There exists three separate cases one needs to check - (i) magnetic field along the Weyl axis of the first parent, $\mathbf{B}=B\hat{y}$, (ii) magnetic field along the Weyl axis of the second parent, $\mathbf{B}=B\hat{\mathbf{z}}$, and (iii) magnetic field perpendicular to the Weyl axis of both parents, $\mathbf{B}=B\hat{\mathbf{x}}$.
\begin{itemize}
\item Case 1 ($\mathbf{B}=B\hat{\mathbf{y}}$) : Substituting, $k_x\rightarrow k_x^\prime =k_x+eBz$, and using the bosonic ladder operators, $a_{\perp,y} = \frac{k_z-ik_x^\prime}{\sqrt{2eB}}$, $a^\dagger_{\perp,y}=\frac{k_z+ik_x^\prime}{\sqrt{2eB}}$, we have, from Eqn. \ref{MWSMperpliny},
\begin{equation}
\begin{split}
H_{\perp,1}(\boldsymbol{k}^\prime) =& (\sin k_{0,1}(k_y-k_{0,1})\tau^z+\sqrt{2eB}(a_{\perp,y}\tau^++a_{\perp,y}^\dagger \tau^-)\\
&\otimes (\sin k_{0,1}\sigma^y+(\gamma_1-\gamma_2)\sigma^z).
\end{split}
\label{MWSMperpBy1}
\end{equation}
For the expression from Eqn. \ref{MWSMperplinz}, we instead consider the following bosonic ladder operators, $\tilde{a}_{\perp,y}=\frac{\tilde{k_z}-ik_x^\prime}{\sqrt{2eB\sin k_{0,2}}}$ and $\tilde{a}_{\perp,y}^\dagger=\frac{\tilde{k_z}+ik_x^\prime}{\sqrt{2eB\sin k_{0,2}}}$, which gives us,
\begin{equation}
\begin{split}
H_{\perp,2}(\boldsymbol{k}^\prime) =& (\sin k_{0,2}\tau^y+(\gamma_1-\gamma_2)\tau^z)\\
&\otimes (k_y\sigma^y-\sqrt{2eB\sin k_{0,2}}(\tilde{a}_{\perp,y}\sigma^+_y+\tilde{a}_{\perp,y}^\dagger \sigma^-_y)).
\end{split}
\label{MWSMperpBy2}
\end{equation}
It is easy to find the lowest Landau level energies in the vicinity of each node. From Eqn. \ref{MWSMperpBy1} and \ref{MWSMperpBy2}, we respectively have the LLL energies,
\begin{equation}
\begin{split}
E_{y,1,LLL} =& \pm\sqrt{\sin^2 k_{0,1}+(\gamma_1-\gamma_2)^2}\sin k_{0,1}(k_y-k_{0,1}),\\
E_{y,2,LLL} =& \pm\sqrt{\sin^2 k_{0,2}+(\gamma_1-\gamma_2)^2}k_y.
\end{split}
\end{equation}
We then find two $k_y$-dependent chiral LLLs connecting the nodes of the first parent, while we have two chiral LLLs at $k_y=0$ due to the second parent, as shown in Fig.~\ref{MWSMperpLandaulevels} (a). The following result was expected if one considers the Landau levels for the parents for different directions of the magnetic field discussed in the previous subsection. For the MWSM perpendicular case, the incident magnetic field in this case is both parallel to the Weyl axis of parent 1 and perpendicular to the Weyl axis of parent 2, so that we get both kinds of Landau levels simultaneously.

\item Case 2 ($\mathbf{B}=B\hat{z}$): This produces results similar to Case 1, as shown in Fig.~\ref{MWSMperpLandaulevels} (b). A similar calculation gives us the lowest Landau level energies,
\begin{equation}
\begin{split}
E_{z,1,LLL} =& \pm\sqrt{\sin^2 k_{0,1}+(\gamma_1-\gamma_2)^2}k_z,\\
E_{z,2,LLL} =& \pm\sqrt{\sin^2 k_{0,2}+(\gamma_1-\gamma_2)^2}\sin k_{0,2}(k_z-k_{0,2}).
\end{split}
\end{equation}
\end{itemize}

\section{Discussion and Conclusion}

In this work, we have introduced the previously-unidentified multiplicative topological semimetal phases of matter, distinguished by Bloch Hamiltonians with a symmetry-protected tensor product structure. Parent Bloch Hamiltonians, with either one or both of the parents being topologically non-trivial, may then be combined in the tensor product to realize multiplicative topological semimetal phases inheriting topology from the parent states.

We consider foundational examples of multiplicative topological semimetals, with Bloch Hamiltonians constructed as tensor products of two-band Bloch Hamiltonians, each characterizing a Weyl semimetal phase. These multiplicative topological semimetal phases are protected by a combination of symmetries of class DIII at the level of the child, and each parent Bloch Hamiltonian in class D. Given the great variety of exotic crystalline point group symmetries considered to protect most recently-identified topological semimetal phases, it is remarkable that the symmetry-protection of these multiplicative semimetal phases is relatively simple, and suggests many additional multiplicative semimetal phases may be identified by enforcing these many other symmetries on parent Bloch Hamiltonians.

We first characterize multiplicative topological semimetal phases in the bulk, showing the bulk spectrum of the child Bloch Hamiltonian depends in a multiplicative way on the spectra of the parent Bloch Hamiltonians: each eigenvalue of the child, at a given point in k-space, is a product of eigenvalues, one from each parent. We furthermore consider two different constructions of the multiplicative Weyl semimetal, either for the case of each parent having a pair of Weyl nodes separated along the same axis in $\boldsymbol{k}$-space (parallel construction), or along perpendicular axes in $\boldsymbol{k}$-space (perpendicular construction). For either construction, the multiplicative symmetry-protected structure can then naturally yield nodal degeneracies reminiscent of Dirac nodes or higher-charge Weyl nodes. However, the multiplicative degeneracies are distinguished from these more familiar quasiparticles by distinctive Wannier spectra signatures in the bulk, and exotic bulk-boundary correspondence. Importantly, bulk characterization by Wannier spectra reveals a complex dependence of Berry connection in the child Bloch Hamiltonian on Berry connection of each parent Bloch Hamiltonian, depending on whether the parents are constructed with Weyl nodes separated along the same axis in momentum-space (parallel) or not (perpendicular). Additionally, the connectivity of Fermi arc surface states for the multiplicative Weyl semimetal is far more complex than in standard Dirac or Weyl semimetals, reflecting the underlying dependence of the child topology on the topology of the parents. An especially interesting example is the realization of topologically-protected---yet floating---boundary states.

Response signatures of the multiplicative Weyl semimetal also inherit response signatures of the parents, with the potential for emergent phenomena beyond that of either parent individually. Here, we consider the multiplicative analog of one of the defining response signatures of the Weyl semimetal, the chiral anomaly, finding instead multiple co-existing chiral anomalies graded by the parent degrees of freedom, as well as emergent features in the Landau level structure not inherited from a particular parent. In the case of parents corresponding to effectively the same degree of freedom, the response reduces to a signature reminiscent of a Dirac semimetal. This brings up the possibility of controlled manipulation of particular properties of an electronic system similar to spintronics.

Future work will characterize other signatures of multiplicative topological semimetals anticipated given the extensive characterization of Weyl and Dirac semimetals, particularly optical and non-linear responses given the tremendous interest in the bulk photovoltaic effect in Weyl semimetals, as well as symmetry-protection of more exotic topological quasiparticles, such as multiplicative generalizations of multifold fermions or nodal lines. Given the immense body of work on topological semimetals and the surprising consequences of multiplicative topology for bulk-boundary correspondence, nodal band structure, and Berry phase structure, our introduction of previously-unidentified multiplicative topological semimetals into the literature lays the foundation for considerable future theoretical and experimental study, which will greatly expand and deepen our understanding of topological semimetal phases.

\textbf{Acknowledgements} - We gratefully acknowledge helpful discussions with J.~E.~Moore, I.~A.~Day, D. Varjas and R.~Calderon.

\textbf{Correspondence} - Correspondence and requests for materials should be addressed to A.M.C. (email: cooka@pks.mpg.de).

\bibliography{ref.bib}

\cleardoublepage


\makeatletter
\renewcommand{\theequation}{S\arabic{equation}}
\renewcommand{\thefigure}{S\arabic{figure}}
\renewcommand{\thesection}{S\arabic{section}}
\setcounter{equation}{0}
\setcounter{section}{0}
\onecolumngrid
\begin{center}
  \textbf{\large Supplemental material for ``Multiplicative topological semimetals''}\\[.2cm]
  Adipta Pal$^{1,2}$, Joe H. Winter$^{1,2,3}$, and Ashley M. Cook$^{1,2,*}$\\[.1cm]
  {\itshape ${}^1$Max Planck Institute for Chemical Physics of Solids, Nöthnitzer Strasse 40, 01187 Dresden, Germany\\
  ${}^2$Max Planck Institute for the Physics of Complex Systems, Nöthnitzer Strasse 38, 01187 Dresden, Germany}\\
  ${}^3$SUPA, School of Physics and Astronomy, University of St.\ Andrews, North Haugh, St.\ Andrews KY16 9SS, UK \\
  ${}^*$Electronic address: cooka@pks.mpg.de\\
\end{center}

\section{Wilson loops for multiplicative Weyl semi-metal:}\label{Wilsonloopsup}
 Labelling the parent Hamiltonians as $\mathcal{H}_{p1}=(d_{1x},d_{1y},d_{1z})\cdot \boldsymbol{\tau}$ and $\mathcal{H}_{p2}=(d_{2x},d_{2y},d_{2z})\cdot \boldsymbol{\sigma}$, with eigenvectors, $\{\ket{+_1}\ket{-_1}\}$ and $\{\ket{+_2},\ket{-_2}\}$ respectively, the child Hamiltonian is given by $\mathcal{H}_{c}=\mathcal{H}_{p1}\otimes\mathcal{H}_{p2}^\prime$, where $\mathcal{H}_{p2}^\prime=(-d_{2x},d_{2y},-d_{2z})\cdot\boldsymbol{\sigma}$. The ground state subspace of the child Hamiltonian is then spanned by, $\{\ket{+_1}\ket{-_2}^\prime,\ket{-_1}\ket{+_2}^\prime\}$ $=\{\ket{+_1}\overline{\ket{+_2}},\ket{-_1}\overline{\ket{-_2}}\}$, where $\overline{\ket{\psi}}$ denotes complex conjugation and ' denotes an eigenstate of $\mathcal{H}_{p2}^\prime$. The non-abelian Berry connection is then given as follows:
\begin{equation}
\begin{split}
A_\mu =& i
\begin{pmatrix}
\bra{+_1,\overline{+}_2}\partial_\mu\ket{+_1,\overline{+}_2} & \\
 & \bra{-_1,\overline{-}_2}\partial_\mu\ket{-_1,\overline{-}_2}\\
\end{pmatrix}
=i
\begin{pmatrix}
\bra{+_1}\partial_\mu\ket{+_1} & \\
 & \bra{-_1}\partial_\mu\ket{-_1}\\
\end{pmatrix}
+i
\begin{pmatrix}
\overline{\bra{+_2}}\partial_\mu\overline{\ket{+_2}} & \\
 & \overline{\bra{-_2}}\partial_\mu\overline{\ket{-_2}}\\
\end{pmatrix}
,
\\
=&
\begin{pmatrix}
A_{1,\mu}^+-A_{2,\mu}^+ & \\
 & A_{1,\mu}^--A_{2,\mu}^-\\
\end{pmatrix}
,
\end{split}
\end{equation}
where $A_{j,\mu}^l = i\bra{l_j}\partial_\mu\ket{l_j}$. For Berry connection around a loop in the Brillouin zone, the values of $\mu$ are $\{k_x,k_y,k_z\}$ for a 3d Brillouin Zone. This clearly shows the difference between the parallel multiplicative phases and the perpendicular multiplicative phases. For a 1d BZ, as shown in past work~\cite{cook2022mult}, the connection for parallel MKC is $\mathbf{A}=(A_{1,k_x}-A_{2,k_x},0,0)$, while for the perpendicular MKC it is, $\mathbf{A}=(A_{1,k_x},-A_{2,k_y},0)$. For 2d or 3d parent systems, it then becomes very straightforward to extrapolate this trend such that the Berry connection looks qualitatively like the combination of the parallel and perpendicular MKC connections based on which directions the parents have in common. This is particularly interesting for the case of parallel and perpendicular Multiplicative Chern Insulators(MCIs), where parent CIs are each defined over a 2d BZ, and the parents can share one or two axes. We illustrate the MCI parallel with two parent CIs on the x-y plane. The resultant Berry connection is then $\mathbf{A}=(A_{1,k_x}-A_{2,k_x},A_{1,k_y}-A_{2,k_y},0)$. The MCI perpendicular on the other hand is constructed with one parent in the x-y plane and another in the x-z plane. The resulting Berry connection is then, $\mathbf{A}=(A_{1,k_x}-A_{2,k_x},A_{1,k_y},-A_{2,k_z})$. The MWSM, on the other hand, is a 3d system, so we  instead consider parent Weyl nodes separated along parallel or perpendicular axes in k-space. As explained in the main text, the parallel MWSM has parent Weyl nodes separated along the same axis in k-space (the $k_z$ axis) while the perpendicular MWSM has parent $1$ and parent $2$ Weyl nodes separated along the $k_y$-axis and $k_z$-axis, respectively. The resultant Berry connection is then, $\mathbf{A}=(A_{1,k_x}-A_{2,k_x},A_{1,k_y}-A_{2,k_y},A_{1,k_z}-A_{2,k_z})$.

\section{Calculation for the surface state spectrum of MWSM:}
We write down here the derivation for the surface state energy for the MWSM parallel and MWSM perpendicular Hamiltonians, for the case of open boundary conditions in the $\hat{x}$ direction and periodic boundary conditions in the $\hat{y}$ and $\hat{z}$ directions. First, we briefly specify how such a calculation should be done for the two band Weyl semi-metal.

\subsection{Slab spectra for WSM:}
We start by writing down the WSM Hamiltonian used,
\begin{equation}
\begin{split}
\mc{H}_{WSM}(\boldsymbol{k})=&t_3(2+\gamma-\cos k_x-\cos k_y-\cos k_z)\sigma^z+t_2\sin k_y\sigma^y+t_1\sin k_x\sigma^x,\\
=&t_3(f-\cos k_x)\sigma^z+t_2\sin k_y\sigma^y+t_1\sin k_x\sigma^x,
\end{split}
\end{equation}
where $f=2+\gamma-\cos k_y-\cos k_z$. Surface states decay into the bulk, so for open boundaries in the x-direction, we carry out the transformation, $k_x\rightarrow iq$ for edge states on the left side $(x=0)$, so that,
\begin{equation}\label{WSMslab0}
\mc{H}_{WSM}(iq,k_y,k_z)=t_3(f-\cosh q)\sigma^z+t_2\sin k_y\sigma^y+it_1\sinh q\sigma^x.
\end{equation}
We claim that the determinant derived from the matrix due to the following limit must be zero,
\begin{equation}\label{WSMslab1}
\lim_{q_1\rightarrow q_2}\frac{\mc{H}(iq_1)-\mc{H}(iq_2)}{2\sinh q_-}=-t_3\sinh q_+\sigma^z+it_1\cosh q_+\sigma^x,
\end{equation}
where $q_\pm=\frac{1}{2}(q_1\pm q_2)$. Carrying out the determinant, we get the following two conditions,
\begin{equation}
t_3\sinh q_+=\pm t_1\cosh q_+.
\end{equation}
Choosing the $+$ sign, the RHS in Eqn. \ref{WSMslab1} becomes, $-t_1\cosh q_+(\sigma^z-i\sigma^x)$, so that the null eigenvector derived from it is one of the eigenvectors for the surface spectra,
\begin{equation}
\ket{\psi_+}=\frac{1}{\sqrt{2}}
\begin{pmatrix}
1\\
i
\end{pmatrix}
.
\end{equation}
The energy corresponding to this eigenvector can be found by solving the eigenvalue for the RHS in Eqn. \ref{WSMslab0} with the above eigenvector. This gives us the eigen-energy, $E$, and the equation to determine the eigen-function, for the left boundary
\begin{subequations}
\begin{equation}
E=t_2\sin k_y,
\end{equation}
\begin{equation}
\begin{split}
&(t_3+t_1)e^{-2q}+2fe^{-q}+(t_3-t_1)=0,\\
\implies &e^{-q_{\pm}} = \frac{-f\pm\sqrt{f^2-(t_3^2-t_1^2)}}{(t_3+t_1)},\\
&\Psi_{+}(x,y,z) \sim (e^{-q_+x}-e^{-q_-x})e^{ik_yy+ik_zz}\ket{\psi_+}.
\end{split}
\end{equation}
\end{subequations}
The eigen-function in the last line has the following form based on the boundary condition on the left edge, $\Psi(x=0)=0$. The other edge can be derived similarly by shifting $x\rightarrow L+1-x$ where L is the length of the system along the x-direction.

\subsection{Slab spectra for MWSM parallel:}
We use the same method as in section S2A of the supplementary materials to derive surface states and spectra for the MWSM parallel system. The Hamiltonian is given as follows,
\begin{equation}
\begin{split}
\mc{H}_{MWSM||}(\boldsymbol{k}) = [t_{31}(f_1-\cos k_x)\tau^z+t_{21}\sin k_y\tau^y+t_{11}\sin k_x\tau^x]\otimes [-t_{32}(f_2-\cos k_x)\sigma^z+t_{22}\sin k_y\sigma^y-t_{12}\sin k_x\sigma^x],
\end{split}
\end{equation}
where $f_{1/2} = 2+\gamma_{1/2}-\cos k_y-\cos k_z$. To ease our calculations, we carry out the following rotation on the four band basis, $\tau^z\rightarrow \tau^y$, $\tau^y\rightarrow -\tau^z$ and $\sigma^z\rightarrow -\sigma^y$, $\sigma^y\rightarrow -\sigma^z$. The Hamiltonian then becomes,
\begin{equation}
\begin{split}
\mc{H}_{MWSM||}(\boldsymbol{k}) =& [t_{31}(f_1-\cos k_x)\tau^y-t_{21}\sin k_y\tau^z+t_{11}\sin k_x\tau^x]\otimes [-t_{32}(f_2-\cos k_x)\sigma^y-t_{22}\sin k_y\sigma^z-t_{12}\sin k_x\sigma^x],\\
=& [t_{31}(f_1-\cos k_x)\tau^y+t_{11}\sin k_x\tau^x][-t_{32}(f_2-\cos k_x)\sigma^y-t_{12}\sin k_x\sigma^x]\\
&-t_{21}\sin k_y\tau^z[-t_{32}(f_2-\cos k_x)\sigma^y-t_{12}\sin k_x\sigma^x]-t_{22}\sin k_y[t_{31}(f_1-\cos k_x)\tau^y+t_{11}\sin k_x\tau^x]\sigma^z\\
&+t_{21}t_{22}\sin^2k_y\tau^z\sigma^z.
\end{split}
\end{equation}
Again, without loss of generality, we set $t_{11}=t_{21}=t_{31}=1=t_{32}=t_{22}=t_{12}$. The edge modes on the left edge $(x=0)$, require we carry out the substitution, $k_x\rightarrow iq$, and the Hamiltonian is now,
\begin{equation}\label{MWSMpllslab0}
\begin{split}
\mc{H}_{MWSM||}(iq,k_y,k_z) =& [(f_1-\cosh q)\tau^y+i\sinh q\tau^x][-(f_2-\cosh q)\sigma^y-i\sinh q\sigma^x]\\
&-\sin k_y\tau^z[-(f_2-\cosh q)\sigma^y-i\sinh q\sigma^x]-\sin k_y[(f_1-\cosh q)\tau^y+i\sinh q\tau^x]\sigma^z\\
&+\sin^2k_y\tau^z\sigma^z.
\end{split}
\end{equation}
Carrying out our previous limit on the rotated Hamiltonian above, we get the following matrix,
\begin{equation}\label{MWSMpllslab1}
\begin{split}
\lim_{q_1\rightarrow q_2}\frac{\mc{H}_{MWSM||}(iq_1)-\mc{H}_{MWSM||}(iq_2)}{2\sinh q_-}=
\begin{pmatrix}
0 & i\sin k_y S_+ & -i\sin k_y S_+ & S_+(-(f_1+f_2)+2S_+)\\
i\sin k_y S_- & 0 & -f_1S_-+f_2S_+ & i\sin k_y S_+\\
-i\sin k_y S_- & f_1S_+-f_2S_- & 0 & -i\sin k_yS_+\\
S_-((f_1+f_2)-2S_-) & i\sin k_yS_- & -i\sin k_yS_- & 0 \\
\end{pmatrix}
,
\end{split}
\end{equation}
where $S_\pm = \cosh q_+\pm \sinh q_+$. The determinant of the RHS of Eqn. \ref{MWSMpllslab1} must be zero, i.e., we have the condition,
\begin{equation}
\begin{split}
&S_-S_+[\sin^2k_yS_-(f_1+f_2-2S_+)(f_1-f_2)(S_-+S_+)-\sin^2 k_y S_+(f_1+f_2-2S_-)(f_1+f_2)(S_+-S_-)\\
&-(f_1+f_2-2S_+)(f_1+f_2-2S_-)(f_1S_--f_2S_+)(-f_2S_-+f_1S_+)]=0
\end{split}
\end{equation}
Let us start with the first condition, $S_-=0$. The RHS of Eqn. \ref{MWSMpllslab1} then becomes,
\begin{equation}
\begin{split}
\lim_{q_1\rightarrow q_2}\frac{\mc{H}_{MWSM||}(iq_1)-\mc{H}_{MWSM||}(iq_2)}{2\sinh q_-}=&S_+(-(f_1+f_2)+2S_+)\tau^+\sigma^++f_2S_+\tau^+\sigma^-+f_1S_+\tau^-\sigma^+\\
&+i\sin k_yS_+\tau^z\sigma^+-i\sin k_yS_+\tau^+\sigma^z,
\end{split}
\end{equation}
where $\tau^{\pm}=\frac{1}{2}(\tau^x\pm i\tau^y)$ and $\sigma^\pm =\frac{1}{2}(\sigma^x\pm i\sigma^y)$ are the two level ladder operators. Here, if $\{\ket{+},\ket{-}\}$ are eigen-vectors of $\tau^z$, then $\tau^+\ket{-}=\ket{+}$, $\tau^+\ket{+}=0$, $\tau^-\ket{-}=0$ and $\tau^-\ket{+}=\ket{-}$. Similar relations exist for the $\sigma$ counterpart. $\ket{\psi_1}=\ket{+}\otimes\ket{+}$ is a null eigen-vector to the above expression on the RHS. We solve for the energy eigenvalue first for the special case $\gamma_1=\gamma_2$.Then from $\mc{H}_{MWSM||}(iq,ky,kz)$ in Eqn. \ref{MWSMpllslab0} due to the eigen-vector $\ket{\psi_1}$, we have the energy and the condition,
\begin{subequations}
\begin{equation}
E=\sin^2 k_y;
\end{equation}
\begin{equation}
(f_1-\cosh q+\sinh q)(f_2-\cosh q+\sinh q)=0.
\end{equation}
\end{subequations}

\section{Landau Level repulsion in the MWSM parallel system:}
We start with the MWSM parallel case,
\begin{equation}
\begin{split}
H_{MWSM,||}(\boldsymbol{k}) =& [t_1(2+\gamma_1-\cos k_x-\cos k_y-\cos k_z)\tau^z+t_1^\prime\sin k_y\tau^y+t_1^\prime\sin k_x\tau^x]\\
&\otimes [-t_2(2+\gamma_2-\cos k_x\cos k_y-\cos k_z)\sigma^z+t_2^\prime\sin k_y\sigma^y-t_2^\prime\sin k_x\sigma^x].
\end{split}
\end{equation}
We expand the Bloch Hamiltonian near the z-axis i.e. $\boldsymbol{k}\rightarrow (0,0,k_z)$,
\begin{equation}
\begin{split}
H_{MWSM,||}(\boldsymbol{k})\approx & [t_1(Q_1+\frac{1}{2}(k_x^2+k_y^2))\tau^z+t_1^\prime k_y\tau^y+t_1^\prime k_x\tau^x]\\
&\otimes [-t_2(Q_2+\frac{1}{2}(k_x^2+k_y^2))\sigma^z+t_2^\prime k_y\sigma^y-t_2^\prime k_x\sigma^x],
\end{split}
\end{equation}
where $Q_i=\gamma_i-\cos k_z$ (i=1,2). Expanding only up to second order in momenta, we have,
\begin{equation}
\begin{split}
H_{MWSM,||}(\boldsymbol{k}) \approx & -t_1t_2(Q_1Q_2+(Q_1+Q_2)\frac{1}{2}(k_x^2+k_y^2))\tau^z\sigma^z\\
&-t_1t_2^\prime Q_1\tau^z(k_x\sigma^x-k_y\sigma^x)-t_1^\prime t_2Q_2(k_x\tau^x+k_y\tau^y)\sigma^z\\
&-t_1^\prime t_2^\prime (k_x^2\tau^x\sigma^x-k_y^2\tau^y\sigma^y-\frac{1}{2}(k_xk_y+k_yk_x)\tau^x\sigma^y+\frac{1}{2}(k_xk_y+k_yk_x)\tau^y\sigma^x).
\label{pllLLbeforesub}
\end{split}
\end{equation}

\begin{figure}[h!]
\centering
\includegraphics[scale=0.7]{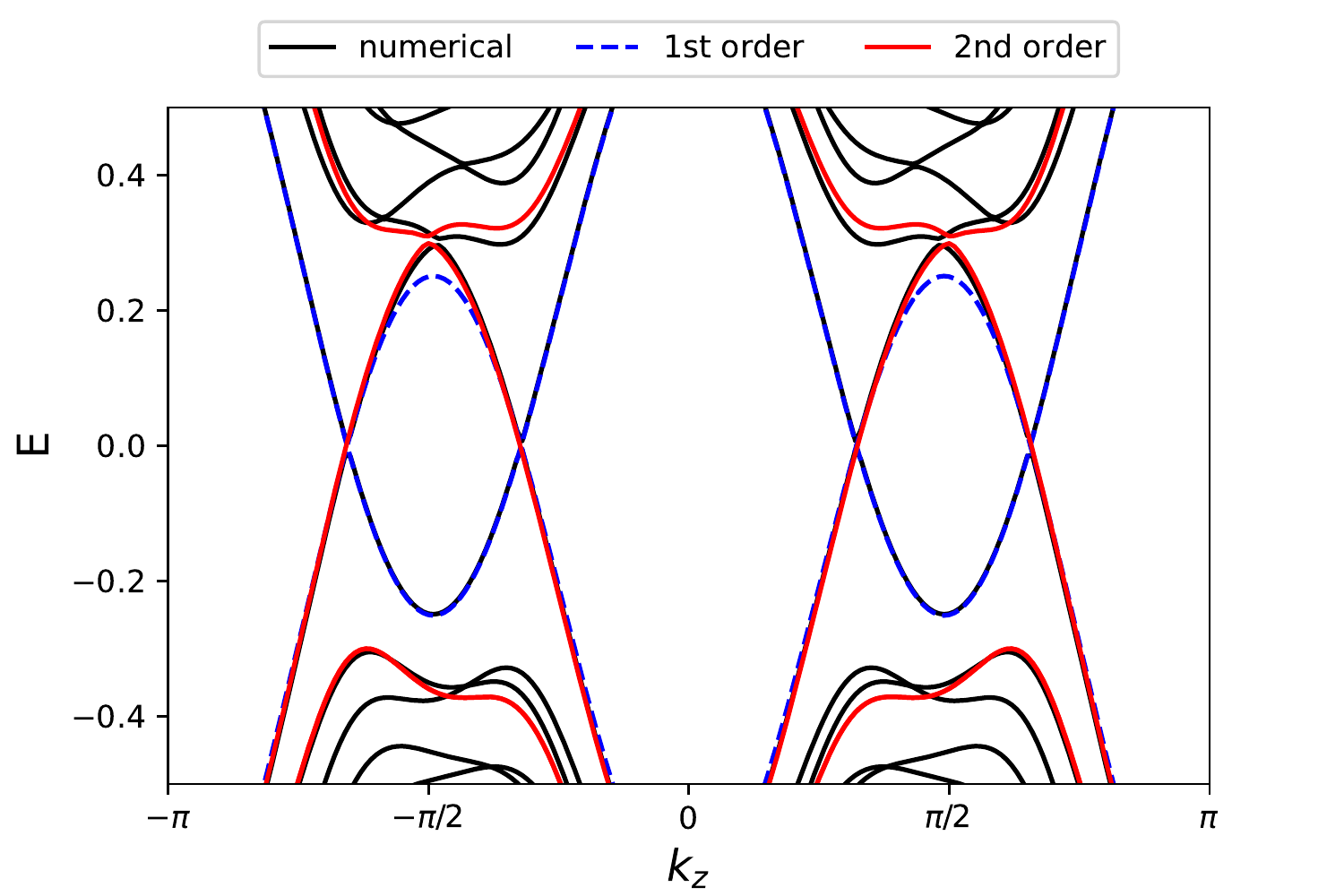}
\caption{Comparison of the numerically calculated Landau Levels of the MWSM$||$. system with the analytically calculated lower Landau levels for first order(blue dashed) and second order(red) expansion in momenta along the direction perpendicular to $k_z$. Level repulsion between two parent graded lowest Landau levels are only observed if one expands to second order in momenta.}
\label{MWSMpllLL2ndorder}
\end{figure}

We consider $\mathbf{B}=B\hat{z}$. After Peierls substitution, $k_x\rightarrow k_x^\prime = k_x$, $k_y\rightarrow k_y^\prime = k_y+eBx$, and $k_z\rightarrow k_z^\prime = k_z$. The position-momenta commutator leads to the commutator, $[k_y^\prime, k_x^\prime] = ieB$. Here, $e$ is the charge of the particle in consideration. One can therefore construct bosonic ladder operators of the form,
\begin{equation}
a = \frac{k_x^\prime -ik_x^\prime}{\sqrt{2eB}};\quad a^\dagger = \frac{k_x^\prime+ik_y^\prime}{\sqrt{2eB}};\quad [a,a^\dagger] = 1.
\label{ktoboson}
\end{equation}
We calculate some important identities via Eqn.\ref{ktoboson} which we will be using in the next few lines,
\begin{equation}
\begin{split}
&\frac{1}{2}({k_x^\prime}^2+{k_y^\prime}^2) = eB(a^\dagger a+\frac{1}{2});\quad k_x^\prime\sigma^x+k_y^\prime\sigma^y = \sqrt{2eB}(a\sigma^++a^\dagger\sigma^-);\quad k_x^\prime\sigma^x-k_y^\prime\sigma^y = \sqrt{2eB}(a\sigma^-+a^\dagger\sigma^+),\\
&{k_x^\prime}^2-{k_y^\prime}^2 = eB(a^2+{a^\dagger}^2);\quad i[k_x^\prime k_y^\prime+k_y^\prime k_x^\prime] = eB({a^\dagger}^2-a^2),
\end{split}
\label{ktobosonrelns}
\end{equation}
where we have used $\tau^\pm = \frac{1}{2}(\tau^x\pm i\tau^y)$ and $\sigma^\pm = \frac{1}{2}(\sigma^x\pm i\sigma^y)$, which are spin ladder operators in the basis $\{\ket{+},\ket{-}\}$ in both the $\tau$ and $\sigma$ spaces. Now, substituting $\boldsymbol{k}$ for $\boldsymbol{k^\prime}$ in Eqn. \ref{pllLLbeforesub} and then transforming them via Eqn. \ref{ktobosonrelns}, we get the following expression,
\begin{equation}
\begin{split}
H_{MWSM,||}(\boldsymbol{k}^\prime)\approx & -t_1t_2(Q_1Q_2+(Q_1+Q_2)eB(a^\dagger a+\frac{1}{2}))\tau^z\sigma^z-t_1t_2^\prime Q_1\sqrt{2eB}\tau^z(a\sigma^-+a^\dagger\sigma^+)-t_1^\prime t_2Q_2\sqrt{2eB}(a\tau^++a^\dagger\tau^-)\sigma^z\\
&-t_1^\prime t_2^\prime (2eB)(a^\dagger a +\frac{1}{2})(\tau^+\sigma^++\tau^-\sigma^-)-t_1^\prime t_2^\prime (2eB)(a^2\tau^+\sigma^-+{a^\dagger}^2\tau^-\sigma^+).
\end{split}
\end{equation}
Let us ignore the second order perturbations not in the mass term (i.e. $\tau^z\sigma^z$) and simplify the Hamiltonian,
\begin{equation}
H_{MWSM,||}(\boldsymbol{k}^\prime) \approx -(Q_1Q_2+(Q_1+Q_2)eB(a^\dagger a+\frac{1}{2}))\tau^z\sigma^z- Q_1\sqrt{2eB}\tau^z(a\sigma^-+a^\dagger\sigma^+)-Q_2\sqrt{2eB}(a\tau^++a^\dagger\tau^-)\sigma^z.
\end{equation}
We obtain one of the lowest Landau levels, $\ket{\psi}_{1,LLL}=\ket{0;-,+}$ with energy $E_{1,LLL} = (Q_1Q_2+\frac{eB}{2}(Q_1+Q_2))$ which match exactly both numerically and analytically in first and second order expansions. For the other lowest Landau level, we observe an amalgamation of chiral Landau levels obtained from each parent which cause level repulsion at the intersection point.

\section{Euler space topology calculation}
In the main text, we have already reported that the MWSM system possesses both time reversal, $\mc{T}$ and inversion symmetry, $\mc{I}$ and hence the combined symmetry, $\mc{T}^\prime$ denoted by $\tau^y\sigma^y\kappa$, where $\kappa$ refers to complex conjugation. However, here ${\mc{T}^\prime}^2=1$, so that a $\mathbb{Z}_2$ invariant is not possible. Instead, it is possible to find a basis, where $\mc{T}^\prime = \kappa$. Here we provide the unitary transformation which makes this possible,
\begin{equation}
V = \frac{1}{2}[(1+i)\tau^0\sigma^0+(1-i)\tau^y\sigma^y].
\end{equation}
Based on the method provided in the appendix in a previous work \cite{guan2022landau}, the above unitary transformation satisfies, $V\tau^y\sigma^yV^T=1$, so that we get a Hamiltonian, $\tilde{H}(\boldsymbol{k})=VH(\boldsymbol{k})V^\dagger$ which satisfies, $\tilde{H}^(\boldsymbol{k})=\tilde{H}^*(\boldsymbol{k})$, and is real and symmetric. Denoting the MWSM in a condensed notation,
\begin{equation}
H = (M_1\tau^z+Q_1\tau^x+R_1\tau^y)\otimes (-M_2\sigma^z-Q_2\sigma^x+R_2\sigma^y),
\end{equation}
we obtain after the transformation,
\begin{equation}
\begin{split}
\tilde{H}=\quad &M_1(-M_2\tau^z\sigma^z-Q_2\tau^z\sigma^x+R_2\tau^x\sigma^0)\\
& -Q_1(M_2\tau^x\sigma^z+Q_2\tau^x\sigma^x+R_2\tau^z\sigma^0)\\
&-R_1(M_2\tau^0\sigma^x-Q_2\tau^0\sigma^z-R_2\tau^y\sigma^y).
\end{split}
\end{equation}
Comparing with the method introduced in \cite{guan2022landau}, it is possible to view the real Hamiltonian as an element of a Real oriented Grassmannian, $\tilde{\mathbb{G}}_{2,4}^\mathbb{R}$ which is diffeomorphic to $S^2\times S^2$. For a given $k_z$, then it is possible to define a mapping from the 2d BZ spanned by $k_x$ and $k_y$ (for MWSM $||$) into $(\mathbf{n_1},\mathbf{n}_2)\in S^2\times S^2$ and the topology of $\tilde{H}$ is then determined by the two skyrmion numbers, $Q[\mathbf{n_1}]=q_1$ and $Q[\mathbf{n_2}]=q_2$ of parent 1 and parent 2, respectively. The Euler class topology is then found from these skyrmion numbers as follows,
\begin{equation}
E_{I}=q_2-q_1;\quad E_{II}=q_2+q_1.
\end{equation}
The Euler numbers are unique up to the mapping $(E_{I},E_{II})\rightarrow (-E_{I},-E_{II})$.

\end{document}